\documentclass[showpacs,a4paper]{revtex4}
\usepackage{indentfirst}
\usepackage[english]{babel}
\usepackage[mathscr]{eucal}
\usepackage[dvips]{graphicx}
\usepackage{amsmath}
\usepackage{subfigure}
\usepackage{caption}

\def\lsim{\mathrel{\raise2pt\hbox to 8pt{\raise -5pt\hbox{$\sim$}\hss{$<$}}}}

\begin{document}
\title{Pseudospin symmetry and the relativistic harmonic oscillator}
\author{R. Lisboa}
\affiliation{Instituto de F{\'\i}sica, Universidade Federal
Fluminense, 24210-340 Niter\'oi, Brazil}
\author{M. Malheiro}
\affiliation{Instituto de F{\'\i}sica, Universidade Federal
Fluminense, 24210-340 Niter\'oi, Brazil}
\author{A. S. de Castro}
\affiliation{Departamento de F{\'\i}sica e Qu{\'\i}mica,
Universidade Estadual Paulista, 12516-410 Guaratinguet\'a, S\~ao
Paulo, Brazil}
\author{P. Alberto}
\affiliation{Departamento de F{\'\i}sica and Centro de F{\'\i}sica
Computacional, Universidade de Coimbra, P-3004-516 Coimbra,
Portugal}
\author{M. Fiolhais}
\affiliation{Departamento de F{\'\i}sica and Centro de F{\'\i}sica
Computacional, Universidade de Coimbra, P-3004-516 Coimbra,
Portugal} 

\begin{abstract}
{A generalized relativistic harmonic oscillator for spin 1/2
particles is studied. The Dirac Hamiltonian contains a scalar
$S$ and a vector $V$ quadratic potentials in the radial
coordinate, as well as a tensor potential
$U$ linear in $r$. Setting either or both combinations $\Sigma=S+V$ and $%
\Delta=V-S$ to zero, analytical solutions for bound states of the
corresponding Dirac equations are found. The eigenenergies and
wave functions are presented and particular cases are discussed,
devoting a special attention to the nonrelativistic limit and the
case $\Sigma=0$, for which pseudospin symmetry is exact. We also
show that the case $U=\Delta=0$ is the most natural generalization
of the nonrelativistic harmonic oscillator. The radial node
structure of the Dirac spinor is studied for several combinations
of harmonic-oscillator potentials, and that study allows us to
explain why nuclear intruder levels cannot be described in the
framework of the relativistic harmonic oscillator in the
pseudospin limit. }
\end{abstract}

\maketitle


\section{\protect Introduction}

\label{Sec:OHR}

The harmonic-oscillator potential for relativistic spin 1/2
particles has received considerable attention by many groups. The
subject is of broad interest since the relativistic harmonic
oscillator plays a role in several areas, namely, in nuclear and
particle physics. In particular, it is the central potential of
the nuclear shell model, and it has also been used as the binding
(and confining) two-body potential for quarks, with applications
in meson and baryon spectroscopy.

A special type of harmonic-oscillator potential is achieved by
replacing the linear momentum operator $\mbox{\boldmath $p$}$ in
the Dirac equation by $\mbox{\boldmath $p$}-\mathrm{i}\,\beta
m\omega \mbox{\boldmath $r$}$, where $\beta$ is the usual Dirac
matrix. This replacement results in a second order differential
equation for the upper and lower components of the Dirac spinor
containing a quadratic potential in the radial coordinate $r$.
Besides the quadratic potential, that second order differential
equation contains a constant spin-orbit term, meaning that, in the
nonrelativistic limit, in which only the upper component
survives, the degeneracy of the energy levels is different from
the one found in the non-relativistic harmonic oscillator. As
explained in the following section, that potential may arise from a
Lorentz tensor interaction in spinor space. In Ref.~\cite{ito} we give
a list of references where this kind of potential was initially
studied. Moshinsky and Szczepaniak \cite{Mosh_S} christened it
\textit{Dirac oscillator} and renewed to a great extent the
interest in the topic. The Dirac oscillator has been applied to
quark confinement and supersymmetry \cite{more} and hadron
spectroscopy \cite{twobody}. Its group symmetries have been
studied in Ref.~\cite{quesne}, its solutions in $2+1$ dimensions and
coupling to magnetic field were worked out in Ref.~\cite{rozmej}, and
finally its wave packets and thermodynamical properties in $1+1$
dimensions were studied in Ref.~\cite{nogami}.

 Another possibility to introduce a harmonic potential in
the Dirac equation is by mixing vector and scalar harmonic
potentials with equal magnitude and sign, aiming at obtaining a
quadratic potential in the Schr\"odinger-like second order
differential equation for each spinor component \cite{smith}.
Kukulin \textit{et al.} \cite{kuku} generalized the problem by
considering a vector-scalar harmonic potential plus the Dirac
oscillator. More recently, there has been a wide interest in
relativistic potentials involving mixtures of vector and scalar
potentials with opposite signs. The interest lies on attempts to
explain the pseudospin symmetry in nuclear physics. Cheng
\textit{et al.} \cite{Chen_Meng}, using a Dirac Hamiltonian with
scalar $S$ and vector $V$ potentials quadratic in space
coordinates, found a harmonic-oscillator-like second order
equation which can be solved analytically for $\Delta =V-S=0$, as
considered before by Kukulin \cite{kuku}, and also for $\Sigma
=S+V=0$. Very recently, Ginocchio solved the triaxial, axial, and
spherical harmonic oscillators for the case $\Delta =0$ and
applied it to the study of antinucleons embedded in nuclei
\cite{gino_oh}. The case $\Sigma=0$ is particularly relevant in
nuclear physics, since it is usually pointed out as a necessary
condition for occurrence of pseudospin symmetry in nuclei \cite
{gino,pmmdm}.

In this work we shall consider a Dirac Hamiltonian that
generalizes the previous ones by simultaneously introducing radial
quadratic potentials for $\Delta$, $\Sigma$ and a linear radial
potential for $U$, the tensor potential defined in the following
section. We will study the cases with $\Delta=0$ and $\Sigma=0$,
for which analytical bound solutions do exist, including also, as
particular cases, the Kukulin potential and the original Dirac
oscillator. The eigenenergies and eigenfunctions are obtained
analytically in the general case. Next we shall analyze particular
cases and pay a special attention to the nonrelativistic limits
and to the case $U=\Sigma=0$, for which pseudospin symmetry is
exact and there are still bound states, as opposed to what happens
with nuclear mean fields. From our analysis we also show that the
case $U=\Delta=0$ is a more natural way to introduce a harmonic
oscillator in the Dirac equation than the usual way with $U\neq 0$
and $\Sigma=\Delta=0$. The eigenenergies have the usual degeneracy
of the nonrelativistic case and the upper component a form
similar to the nonrelativistic wave function. Actually, by
letting the harmonic-oscillator frequency to become small compared
with the mass, we obtain exactly the eigenvalues and wave
functions of the nonrelativistic harmonic oscillator.

We also present the node structure of the radial wave functions,
motivated by a study carried on in Ref.~\cite{gino2} for $V$ and $S$
radial potentials vanishing as $r\to\infty$. In order to find
reasons for the unusual radial node structures in some particular
cases, we obtain the relations between the radial nodes of the
upper and lower components of the Dirac spinors by inspection of
their analytical forms, and illustrate those relations by plots of
several of those components. We draw conclusions regarding the
impossibility to describe the so-called intruder states by
harmonic-oscillator potentials with exact pseudospin symmetry.

 This paper is organized as follows. In Sec.~\ref{Sec:OscGeral} we
present the general Dirac equation with scalar and vector
potentials with harmonic oscillator form, proportional to $r^2$,
and a tensor potential, linear in $r$. We then obtain and discuss
the solutions of this equation for $\Delta=0$ and $\Sigma=0$,
respectively. The Dirac oscillator is presented in Sec.~\ref{SubSec:OscDirac}. 
In Sec.~\ref{SubSec:omega2_0_D_0}
and Sec.~\ref{SubSec:omega2_0_S_0} we look into the particular cases
$\Delta=0$ and $\Sigma=0$ in the absence of the tensor potential.
We analyze the
nonrelativistic limits, showing that such a limit does not exist in the case $%
\Sigma=0$, up to first order in the oscillator frequency divided
by the mass. This result is connected to the pseudospin symmetry,
as discussed in more detail in Sec.~\ref{PSymmetry}. In the
same section we also present an analysis of the intruder states
and show the impossibility to describe them within the harmonic
oscillator in the case of exact pseudospin symmetry. Finally, our
conclusions are summarized in Sec.~\ref{conclusions}.


\section{\protect Generalized relativistic harmonic oscillator}

\label{Sec:OscGeral}

 The time-independent Dirac equation for a spin $1/2$ fermion with
energy $\mathcal{E}$, in the presence of a potential, reads
\begin{equation}
H_{\mathrm{D}}\Psi =\mathcal{E} \Psi \ ,  \label{1}
\end{equation}
where the Dirac Hamiltonian is
\begin{equation}
H_{\mathrm{D}}=\mbox{\boldmath $\alpha $}\cdot \mbox{\boldmath $p$}+\beta m+%
\mathcal{V} \, .  \label{2}
\end{equation}
In this Hamiltonian, $m$ is the fermion mass, $\mbox{\boldmath $p$} $ is the
momentum operator, and $\mbox{\boldmath $\alpha $}$ and $\beta $ are $4\times
4 $ matrices which, in the usual representation, take the form
\begin{equation}
\mbox{\boldmath $\alpha $}=\left(
\begin{array}{cc}
0 & \mbox{\boldmath $\sigma $} \\
\mbox{\boldmath $\sigma $} & 0
\end{array}
\right), \qquad \beta = \left(
\begin{array}{cc}
I & 0 \\
0 & -I
\end{array}
\right)\, .
\end{equation}
Here $\mbox{\boldmath $\sigma $}$ is a three-vector whose components are the
Pauli matrices, and $I$ stands for the $2\times 2$ identity matrix. The
matrix potential $\mathcal{V}$ in Eq.~(\ref{2}) may, in general, be written
as a linear combination of sixteen linearly independent matrices, classified
according to their properties under Lorentz transformations: scalar,
pseudoscalar, vector, pseudovector, and tensor. In the following, we exclude
the pseudoscalar and pseudovector potentials, only considering spherically
symmetric potentials, for which the total angular momentum of the system is
constant. In spite of being called \emph{spherically symmetric potentials},
they do not necessarily depend only on the radial coordinate $r=|%
\mbox{\boldmath $r$}|$, because of their matrix structure. Indeed, the
following potential
\begin{equation}
\mathcal{V}(\mbox{\boldmath $r$})=V(r)+\beta S(r)+i\beta
\mbox{\boldmath $\alpha
$}\cdot \mbox{\boldmath $\hat{r}$}\,U(r)  \label{3}
\end{equation}
is spherically symmetric in this sense, because it commutes with the total
angular momentum $\mbox{\boldmath $J$}=\mbox{\boldmath $L$}+%
\mbox{\boldmath $S$}$, where $\mbox{\boldmath $L$}$ and
$\mbox{\boldmath $S$}$ are the orbital and spin angular momentum
operators, respectively. In the last term, $\mbox{\boldmath
$\hat{r}$}=\mbox{\boldmath $r$}/r$, and the radial functions in
Eq.~(\ref{3}) are named after the properties their respective
terms have under Lorentz transformations: $V$ corresponds to the
time component of a vector potential, $S$ is a scalar potential,
and $U$ is a tensor potential. It is important to point out that
the Dirac equation for the potential in Eq.~(\ref{3}) is invariant
under spatial inversion, hence the eigenstates have definite
parity.

It is worth showing where the tensor character of the last term in
Eq.~(\ref{3}) comes from. The interaction Lagrangian for the
tensor coupling of a Dirac spinor $\Psi$ with an external field is
\begin{equation}
\frac{f}{4M}\,\bar\Psi\sigma^{\mu\nu}\Psi\phi_{\mu\nu} \ ,
\end{equation}
where $\phi_{\mu\nu}=\partial_\mu\phi_\nu-\partial_\nu\phi_\mu$, $%
\sigma^{\mu\nu}=(i/2)[\gamma^\mu,\gamma^\nu]$, and $%
\gamma^\nu $ are the Dirac matrices. Considering only the time
component of the vector potential $\phi^\mu$, $\phi^0$, the
corresponding term leads to the following potential in the
single-particle (Dirac) Hamiltonian:
\begin{equation}
i\,\frac{f}{2M}\,\beta \mbox{\boldmath $\alpha
$}\cdot\nabla\phi^0 \ .
\end{equation}
For a static radial potential $\phi^0$, the tensor potential
$U(r)$ in Eq.~(\ref{3}) is $U(r)=f/(2M)\,(d\phi^0/
dr)$. This term can be used to describe a particle with an
anomalous magnetic moment. Indeed, the anomalous magnetic
interaction has the form $-i\mu \beta \mbox{\boldmath $\alpha$}%
\cdot \nabla\varphi (r)$, where $\mu $ is the anomalous magnetic moment, in
units of Bohr magnetons, and $\varphi $ is the electric potential,
i.e., the time component of the electromagnetic vector potential, $A^\mu$.
Therefore, the interaction $-i m\omega \beta
\mbox{\boldmath
$\alpha$}\cdot\mbox{\boldmath $r$}$ (notice that $\beta$ and $%
\mbox{\boldmath $\alpha$}$ are now written in reverse order),
which defines the ``Dirac oscillator" mentioned above, could be
conceived as an anomalous magnetic interaction generated by an
electrostatic harmonic potential.

 In summary, for this type of potentials, the Hamiltonian, the parity
operator, and the total angular momentum form a complete set of commuting
observables. Accordingly, the eigenstates can be classified by the parity ($%
\pm $), by the total angular momentum $j$, and its third
component $m$, quantum numbers. Under these circumstances it is
natural to use spherical coordinates and the spinor, which is the
solution of the Dirac equation (\ref {1}), can be generally
written as
\begin{equation}
\Psi_{jm}^{\pm }(\mbox{\boldmath $r$})=\left(
\begin{array}{c}
\displaystyle i\frac{g_{j\,l}^{\pm }(r)}{r} \mathcal{Y}_{jm}^{\pm }(%
\mbox{\boldmath $\hat{r}$}) \\[.2cm]
\displaystyle\frac{f_{j\,\tilde l}^{\pm }(r)}{r} \mathcal{Y}_{jm}^{\mp }(%
\mbox{\boldmath $\hat{r}$})
\end{array}
\right)\, .  \label{4}
\end{equation}
In this expression, $\mathcal{Y}_{jm}$ are the so-called spinor
spherical harmonics. They result from the coupling of the
two-dimensional spinors to the eigenstates of orbital angular
momentum and form a complete orthonormal set. The orbital angular
momentum quantum numbers $l$ and $\tilde l$ refer to the upper and
lower components, respectively. The normalization of the Dirac
spinor in Eq.~(\ref{4}) implies the following normalization
condition on the upper and the lower radial functions:
\begin{equation}
\int_{0}^{\infty }( |g_{j\,l}^{\pm }|^{2}+|f_{j\,\tilde l}^{\pm
}|^{2}) dr=1\, ,  \label{4a}
\end{equation}
so that $g_{j\,l}^{\pm }$ and $f_{j\,\tilde l}^{\pm }$ should be
square-integrable functions.

Using the operator $\mbox{\boldmath $\alpha$}\cdot%
\mbox{\boldmath
$p$}\, $ in the spherical-polar form, Eq.~(\ref{1}) can be brought to the
form
\begin{equation}
\biggl\{-i\mbox{\boldmath $\alpha$}\cdot\mbox{\boldmath $\hat{r}$}%
\biggl[\frac{\partial}{\partial\,r}+ \frac{1}{r}(1+\beta\,K)\biggr]+\beta[%
m+S(r)]+V(r)+i\,\beta\mbox{\boldmath $\alpha$}\cdot%
\mbox{\boldmath
$\hat{r}$}\, U(r)\biggr\}\Psi=\mathcal{E}\Psi\,,  \label{Eq:OscGerPolar}
\end{equation}
where we have introduced the operator $K=-\beta(\mbox{\boldmath $\Sigma$}\,%
\cdot\mbox{\boldmath $L$}+1)$ whose eigenvalues are
$\kappa=\pm(j+1/2)$. Here $\mbox{\boldmath $\Sigma$}$ refers to
the $4\times 4$ matrix whose block diagonal elements are the Pauli
matrices, $\mbox{\boldmath $\sigma$}$, and whose off-diagonal
blocks are zero. More explicitly, the spin-orbit coupling quantum
number $\kappa$ is related to the orbital angular momentum quantum
number by
\begin{equation}
\kappa =\left\{
\begin{array}{ccc}
-(l+1) & =-\left( j+1/2\right) , & j=l+1/2\ \ \ (\kappa<0)\; \\[0.2cm]
l & =+\left( j+1/2\right) , & j=l-1/2\ \ \ (\kappa>0)\;
\end{array}
\ \ \
\begin{array}{l}
\text{\textrm{aligned spin}} \\
\text{\textrm{unaligned spin}}.
\end{array}
\right.  \label{5}
\end{equation}
Actually, the quantum number $\kappa$ completely determines $j$, $l$, $%
\tilde l$ and, thus, the parity $(-1)^l$:
\begin{eqnarray}
j&=&|\kappa|-\frac12\,,  \nonumber \\
l&=&|\kappa|+\frac12\Big(\frac{\kappa}{|\kappa|}-1\Big)\,,  \nonumber \\
\tilde l&=&l-\frac{\kappa}{|\kappa|} \, . \label{l_tilde}
\end{eqnarray}
Notice that $\tilde l$ is given by the same formula as $l$, but with $%
-\kappa $ instead of $\kappa$. Accordingly, the Dirac spinors in
Eq.~(\ref{4}) can be labeled just with $\kappa$ and $m$,
i.e.,
\begin{equation}
\Psi_{\kappa m}(\mbox{\boldmath $r$})=\left(
\begin{array}{c}
\displaystyle i\frac{g_{\kappa}(r)}{r} \mathcal{Y}_{\kappa m}(%
\mbox{\boldmath $\hat{r}$}) \\
\noalign{\vskip.2cm} \displaystyle \frac{f_\kappa(r)}{r} \mathcal{Y}%
_{-\kappa m}(\mbox{\boldmath $\hat{r}$})
\end{array}
\right) \, .  \label{6}
\end{equation}
Using the property $\mbox{\boldmath $\sigma $}\cdot
\mbox{\boldmath $\hat{r}$}\,\mathcal{Y} _{\kappa
m}=-\mathcal{Y}_{-\kappa m}$, the Dirac equation in Eq.~(\ref{1})
may be reduced to a set of two coupled first order
ordinary differential equations for the radial upper and lower components $%
g_{\kappa}$ and $f_{\kappa}$, namely,
\begin{eqnarray}
\biggl[\frac{d \ }{d r}+\frac{\kappa}{r}+U(r)\biggr]%
g_{\kappa}(r)= &[\mathcal{E}+m-\Delta(r)]f_{\kappa}(r)\,,  \label{Eq:D1ordRadup}
\\[0.5cm]
\biggl[\frac{d \ }{d r}-\frac{\kappa}{r}-U(r)\biggr]%
f_{\kappa}(r)= &-[\mathcal{E}-m-\Sigma(r)]g_{\kappa}(r)\, ,
\label{Eq:D1ordRadlow}
\end{eqnarray}
where we have introduced the ``sum" and the ``difference" potentials defined
by
\begin{equation}
\Sigma=V+S \ \ \ \mbox{\rm and} \ \ \ \Delta = V-S\, .
\end{equation}

 Using the expression for $f_{\kappa}$ obtained from Eq.~(\ref{Eq:D1ordRadup})
and inserting it in Eq.~(\ref{Eq:D1ordRadlow}) we arrive at
the following second order differential equation for $g_{\kappa}$:
\begin{eqnarray}
\biggl\{\frac{d^{2}\,}{dr^{2}} &-&\frac{\kappa (\kappa +1)%
}{r^{2}}+\frac{\Delta ^{\prime }}{\mathcal{E}+m-\Delta (r)}\biggl[\frac{%
d\,\,}{dr}+\frac{\kappa }{r}+U(r)\biggr]-2\kappa \frac{U(r)%
}{r}+U^{\prime }(r)-U^{2}(r)\biggr\}g_{\kappa }(r)  \nonumber \\
 &=&-[\mathcal{E}-m-\Sigma (r)][\mathcal{E}+m-\Delta (r)]g_{\kappa }(r)\,,
\label{Eq:D2ordgOHGeral}
\end{eqnarray}
where the prime means derivative with respect to $r$. In a
similar fashion, i.e., using again both
Eqs.~(\ref{Eq:D1ordRadup}) and (\ref {Eq:D1ordRadlow}), a second
order differential equation for the lower component is obtained:
\begin{eqnarray}
\biggl\{\frac{d^{2}\,}{dr^{2}} &-&\frac{\kappa (\kappa -1)%
}{r^{2}}+\frac{\Sigma ^{\prime }}{\mathcal{E}-m-\Sigma (r)}\biggl[\frac{%
d\,}{dr}-\frac{\kappa }{r}-U(r)\biggr]-2\kappa \frac{U(r)}{%
r}-U^{\prime }(r)-U^{2}(r)\biggr\}f_{\kappa }(r)=  \nonumber \\
&=&-[\mathcal{E}-m-\Sigma (r)][\mathcal{E}+m-\Delta (r)]f_{\kappa }(r).
\label{Eq:D2ordfOHGeral}
\end{eqnarray}

 In this paper we shall consider harmonic-oscillator potentials,
meaning that potentials $\Sigma $ and $\Delta $ are quadratic in
$r$ and the potential $U$ linear in $r$. With all these potentials
in place, Eqs.~(\ref {Eq:D2ordgOHGeral}) and
(\ref{Eq:D2ordfOHGeral}) have to be solved numerically because of
the quartic potentials in $r$ arising from the product
$[\mathcal{E}-m-\Sigma (r)][\mathcal{E}+m-\Delta (r)]$ and the
terms with derivatives of $\Delta$ and $\Sigma$. However, either
for $\Delta =0$ or for $\Sigma =0$ the solutions are analytical
since, in the first case, there are only quadratic potentials in
the second order equation for $g_\kappa$
[Eq.~(\ref{Eq:D2ordgOHGeral})], and, in the second case, the same
happens in Eq.~(\ref{Eq:D2ordfOHGeral}) for $f_\kappa$. The case
$\Delta=U=0$ allows us to get, in the nonrelativistic limit, the
spectrum  and wave functions of the central nonrelativistic
harmonic oscillator, i.e, Eq.~(\ref{Eq:D2ordgOHGeral})
becomes the radial Schr\"odinger equation for the
three-dimensional harmonic oscillator. For $U=0$, both cases $\Sigma =0$ and $%
\Delta =0$ correspond to SU(2) symmetries of the Dirac equation
\cite {bell,levi}. The case $\Sigma =0$ is related to the
pseudospin symmetry in nuclei.

 The aim of this paper is to study such analytical solutions and, in particular,
to draw conclusions on the required conditions for the pseudospin
symmetry to show up.

 Let us first consider
\begin{equation}
\Sigma (r)=\frac{1}{2}m\,\omega _{1}^{2}\,r^{2}\,,\ \ \ \Delta (r)=0\,,\ \ \
U(r)=m\,\omega _{2}\,r\,,  \label{potenx}
\end{equation}
where $\omega _{2}$ (a real number) is the frequency related to
the tensor potential $U$ and $\omega_{1}$ (a non-negative number
\footnote{In fact, this condition can be relaxed to demand that
$\omega_1$ be just a real number, since the results do not depend
on the sign of $\omega_1$.}) the frequency related to the ``sum''
potential $\Sigma $. Actually, this case was already considered
by  Kukulin \textit{et al.} \cite{kuku} for $\omega _{2}>0$. The
possibility of Dirac bound states with a linear tensor potential,
which can be unbounded from below, is due to the appearance of
$U^{2}$ in Eqs.~(\ref{Eq:D2ordgOHGeral}) and (\ref
{Eq:D2ordfOHGeral}). It is noteworthy that the different signs of
$\omega _{2}$ give rise to different possibilities of signs for
the spin-orbit coupling. Equation (\ref{Eq:D2ordgOHGeral}) for the
upper component takes the form
\begin{equation}
\biggl\{\frac{d^{2}\,}{dr^{2}}-\frac{\kappa (\kappa +1)}{%
r^{2}}-\biggl[\frac{1}{2}m\omega _{1}^{2}(\mathcal{E}+m)+m^{2}\omega _{2}^{2}%
\biggr]r^{2}-(2\kappa -1)m\,\omega _{2}+(\mathcal{E}^{2}-m^{2})\biggr\}%
g_{\kappa }(r)=0\,.  \label{ggg}
\end{equation}
It is convenient to introduce the following new variable and parameters:
\begin{eqnarray}
y &=&\sqrt{\frac{m(\mathcal{E}+m)}{2}}\,\Omega \,r^{2}=a^{2}\,r^{2}\,,
\label{xx1} \\[0.5cm]
\Omega  &=&\sqrt{\frac{2m}{\mathcal{E}+m}\omega _{2}^{2}+\omega _{1}^{2}}\,, \\%
[0.5cm]
\lambda  &=&\frac{(2\kappa -1)m\omega _{2}-(\mathcal{E}^{2}-m^{2})}{a^{2}}\,,
\label{xx3}
\end{eqnarray}
which allows us to write Eq.~(\ref{ggg}) in the simpler form
\begin{equation}
\biggl\{4y\frac{d^{2}\ }{dy^{2}}+2\frac{d\ }{%
dy}-\frac{l(l+1)}{y}-y-\lambda \biggr\}g_{\kappa }(y)=0,
\label{Eq:D2ordg(y)}
\end{equation}
since $\kappa (\kappa +1)=l(l+1)$. An asymptotic analysis suggests searching
for solutions of the type
\begin{equation}
g_{\kappa }(y)=A\,e^{-y/2}\,y^{(l+1)/2}\,\Gamma (y)  \label{Eq:g(y)}
\end{equation}
where $\Gamma (y)$ is a function yet to be determined and $A$ a
normalization constant determined by Eq.~(\ref{4a}). Inserting
this expression back into Eq.~(\ref{Eq:D2ordg(y)}) the equation
for $\Gamma (y)$ reads
\begin{equation}
\biggl[y\frac{\mathrm{d}^{2}\ }{\mathrm{d}y^{2}}+\biggl(l+\frac{3}{2}-y%
\biggr)\frac{\mathrm{d}\ }{\mathrm{d}y}-\frac{1}{2}\biggl(l+\frac{3}{2}+%
\frac{\lambda }{2}\biggr)\biggr]\Gamma (y)=0\,.  \label{Eq:D2ordGamma(y)}
\end{equation}
The solutions of this equation, which guarantee that $\lim_{y\to
\infty}g_{\kappa }(y)=0$, are the generalized Laguerre polynomials of degree $n$, $%
L_{n}^{p}(y)$, where
\begin{eqnarray}
n &=&-\frac{1}{2}\biggl(l+\frac{3}{2}+\frac{\lambda }{2}\biggr)\,,
\label{cond_lambda} \\
p &=&l+\frac{1}{2}\,.
\end{eqnarray}

 From Eqs.~(\ref{cond_lambda}) and (\ref{xx1})--(\ref{xx3}), we can get immediately the
eigenenergies, which are discrete since $n$ is an integer greater
than or equal to zero:
\begin{equation}
\mathcal{E}^{2}-m^{2}-(2\kappa -1)m\omega _{2}=\left( 2n+l+\frac{3}{2}%
\right) \sqrt{2m(\mathcal{E}+m)\omega _{1}^{2}+4m^{2}\omega _{2}^{2}}\, .
\label{Eq:AutD0}
\end{equation}
If we square Eq.~(\ref{Eq:AutD0}) the resulting algebraic equation
is quartic in $\mathcal{E}$. For particular values of $n$ and
$\kappa$ (remember that $\kappa$ determines uniquely $l$) we can
solve this equation with respect to $\mathcal{E}$ and get the
energy of the level with quantum numbers $(n,\kappa )$ or, using
the standard spectroscopic notation, the level $n\,l_{j}$. With
the help of the program MATHEMATICA we were able to get analytical
expressions for those solutions. In general, it can be shown that
there are at most two real bound state solutions of
Eq.~(\ref{Eq:AutD0}), one with a positive or zero binding energy
and the other one with a negative binding energy. As we will see
later, it can happen that only positive binding energy solutions
exist. When both solutions exist, we consider the one with
positive binding energy, i.e., for which $\mathcal{E}>m$.

\begin{figure}[!ht]
\begin{center}
\includegraphics[width=9cm]{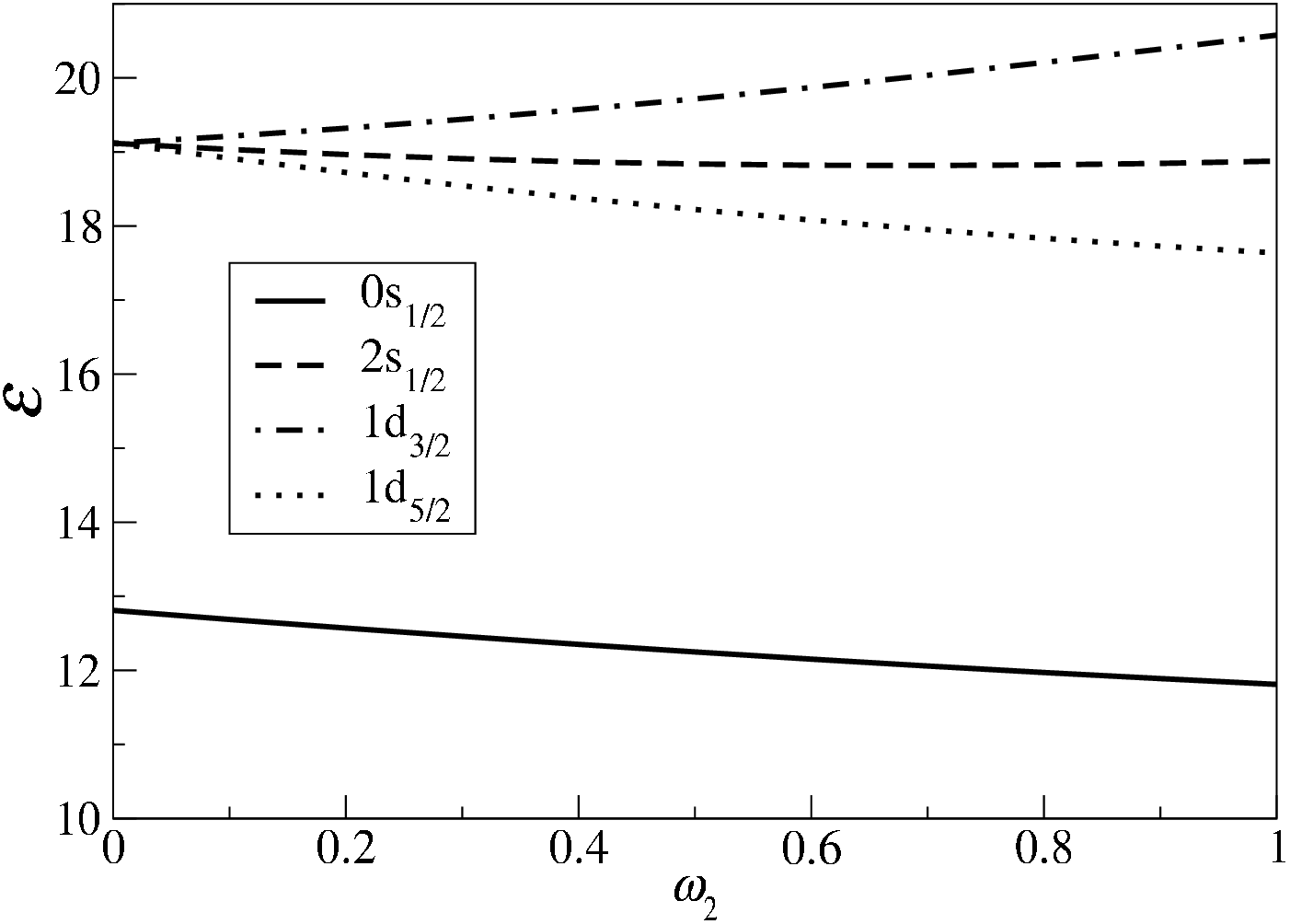}
\end{center}
\par
\vspace*{-0.4cm} \caption{Energy levels for $\Delta =0$ with
$\omega _{1}=2$ and $m=10$ as a function of $\omega _{2}$.}
\label{Fig:enDel}
\end{figure}

\begin{figure}[!ht]
\begin{center}
\includegraphics[width=9cm]{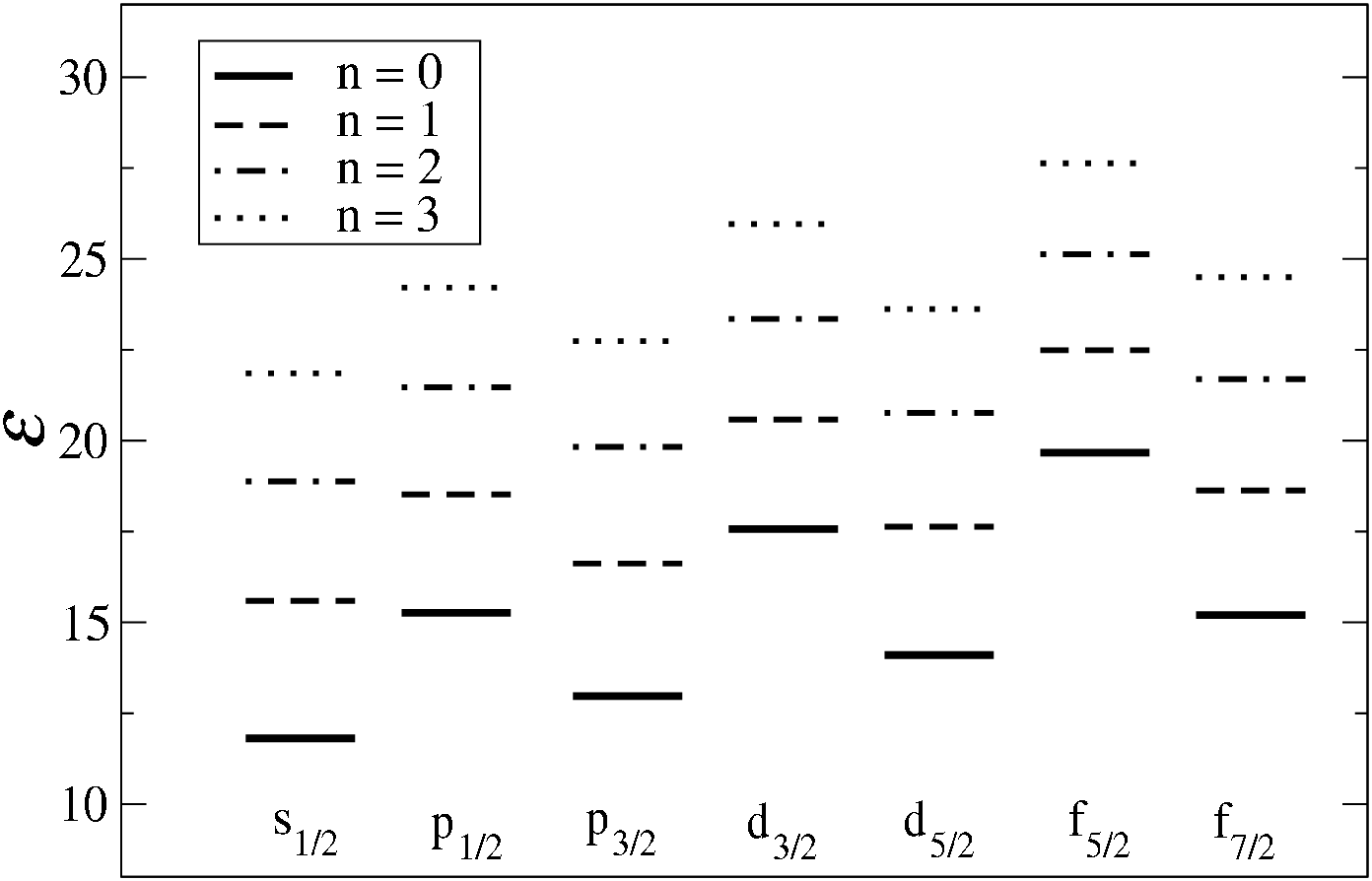}
\end{center}
\par
\vspace*{-0.1cm} \caption{Energy spectrum for $\Delta=0$ with
$\omega_{1}=2$, $\omega_{2}=1$ and $m=10$.} \label{Fig:spcDel}
\end{figure}

Figure {\ref{Fig:enDel}} displays the energy for some levels, as a
function of $\omega_{2}$, for a fixed $\omega_{1}$ and $m$. From
this figure we see that for $\omega_{2}=0$ the energy levels with
$(n=1\,,l=2)$ are degenerate with the level with $(n=2\,,l=0)$ and
have no spin-orbit splitting. This is a manifestation of the
$(2n+l)$ degeneracy of the levels which can be inferred from
Eq.~(\ref{Eq:AutD0}) when $\omega_2$ is set to zero. We will
further elaborate on this point in the following section.
The spectrum for the first seven levels for $\omega_{2}=2$ and
$\omega_{1}=1$ is shown in Fig.~{\ref{Fig:spcDel}}, where we see
that they are, in general, nondegenerate.
This feature of the spectrum can be understood by looking at
Eq.~(\ref{Eq:AutD0}), since when the tensor potential
is present ($\omega_{2}\neq 0$) we have always a
$\kappa$-dependent term (a spin-orbit term) that removes the
$(2n+l)$ degeneracy, which is also characteristic of the
nonrelativistic harmonic-oscillator spectrum.

 It is instructive to get the nonrelativistic limit of Eq.~(\ref
{Eq:AutD0}). We obtain this limit by letting $\omega_1/m$ and $\omega_2/m$
become very small. In this limit, $(\mathcal{E}-m)/m=E/m$ also becomes very
small, and $\mathcal{E}+m\sim 2m$, so that
\begin{equation}
E=\left( 2n+l+\frac{3}{2} \right)\sqrt{\omega_1^2+\omega_2^2}+\bigg(%
\kappa-\frac12\bigg)\omega_2 \ .  \label{Eq:E-non-rel-Delta_0}
\end{equation}

 The solution of Eq.~(\ref{ggg}), replacing $y$ by $a^{2}\, r^{2}$, is
\begin{equation}
g_\kappa(r)=A\;\exp \Bigr(\!-\frac{1}{2}\,a^2\,r^{2}\Bigl)\ (a^2\,r^{2})^{%
(l+1)/2}\ L^{l+1/2}_{n} (a^2\,r^{2})\, .  \label{Eq:Grfull}
\end{equation}
For the lower component, the first order differential equation in
Eq.~(\ref {Eq:D1ordRadup}) allows us to write
\begin{equation}
f_\kappa(r)=\frac{1}{\mathcal{E}+m}\biggr[\frac{d \ }{d r}+%
\frac{\kappa}{r}+ m \omega_2\, r \biggl]g_\kappa(r).  \label{Eq:Ffull}
\end{equation}
Using the recursion relations for the generalized Laguerre
polynomials (see, for example, Ref.~\cite{abramowitz}) we get, for
$\kappa<0$,
\begin{equation}  \label{Eq:Gen_f_kappa_neg}
f_\kappa(r)=\frac{A\,a}{\mathcal{E}+m}\;\exp \Big(\!-\frac{1}{2}\,a^2\,r^{2}%
\Big) \big(a^2\,r^{2}\big)^{(l+2)/2} \bigg[ \bigg(\frac{m\omega_2}{a^2}+1%
\bigg) L^{l+1/2}_{n} (a^2\,r^{2})
-2\,L^{l+3/2}_{n} (a^2\,r^{2})\bigg]
\end{equation}
and, for $\kappa>0$,
\begin{eqnarray}  \label{Eq:Gen_f_kappa_pos}
f_\kappa(r)&=&\frac{A\, a}{\mathcal{E}+m}\;\exp \Big(\!-\frac{1}{2}\,a^2\,r^{2}%
\Big)\ \big(a^2\,r^{2}\big)^{l/2}\times  \nonumber \\
&& \bigg[ \bigg(n+l+\frac12\bigg)\bigg(1+\frac{m\omega_2}{a^2}\bigg)\,L^{l-%
1/2}_{n} (a^2\,r^{2}) +(n+1)\bigg(1-\frac{m\omega_2}{a^2}\bigg)%
\,L^{l-1/2}_{n+1} (a^2\,r^{2})\bigg]\, .
\end{eqnarray}

\begin{figure}[!hb]
{
\parbox[!hb]{5.5cm}{
\begin{center}
\includegraphics[width=5.5cm,height=5.5cm]{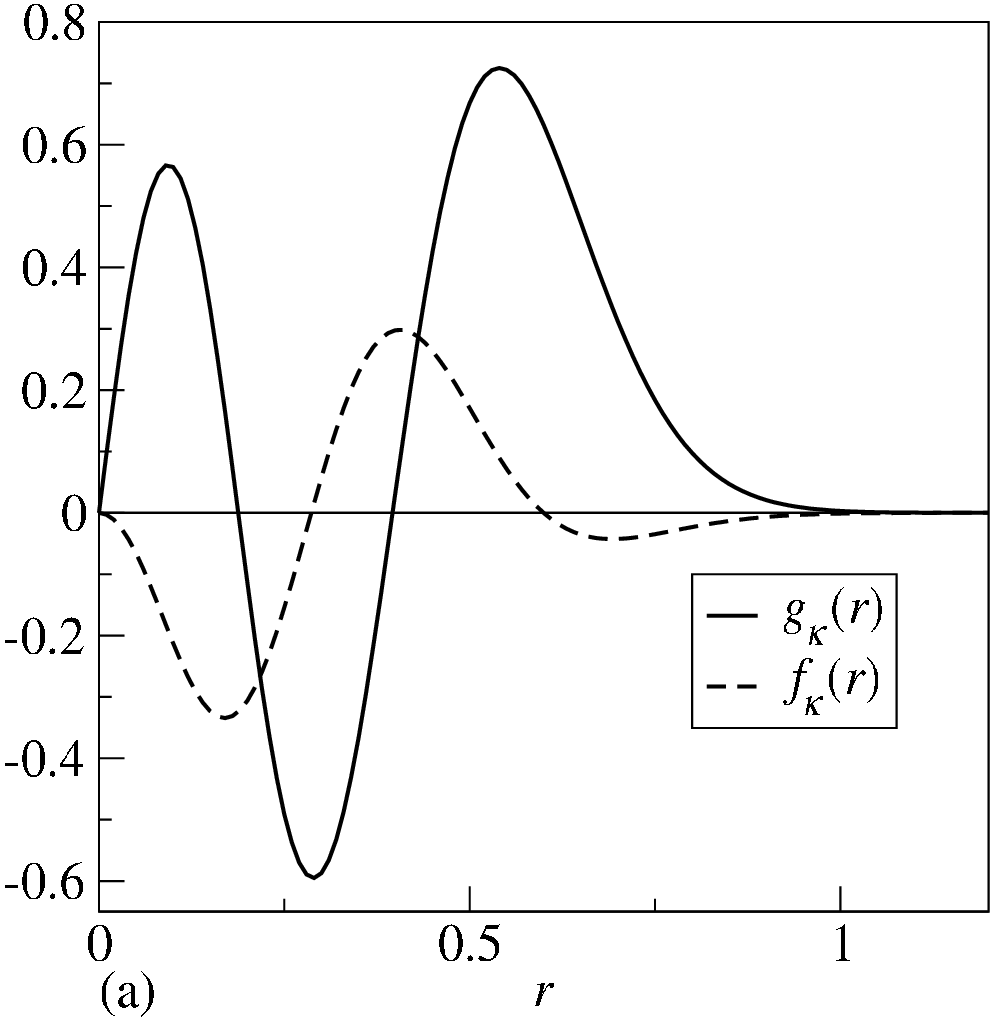}
\end{center}}
\parbox[!hb]{5.5cm}{
\begin{center}
\includegraphics[width=5.5cm,height=5.5cm]{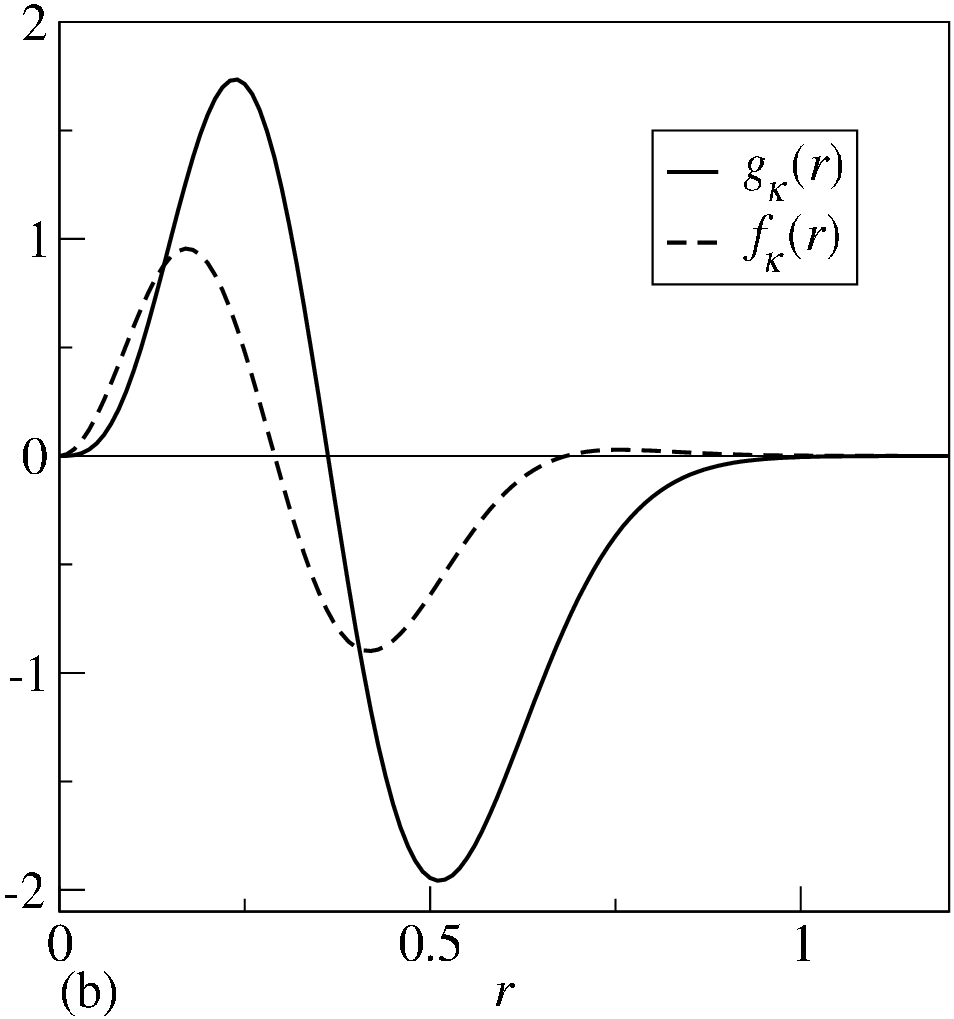}
\end{center}}
\parbox[!hb]{5.5cm}{
\begin{center}
\includegraphics[width=5.5cm,height=5.5cm]{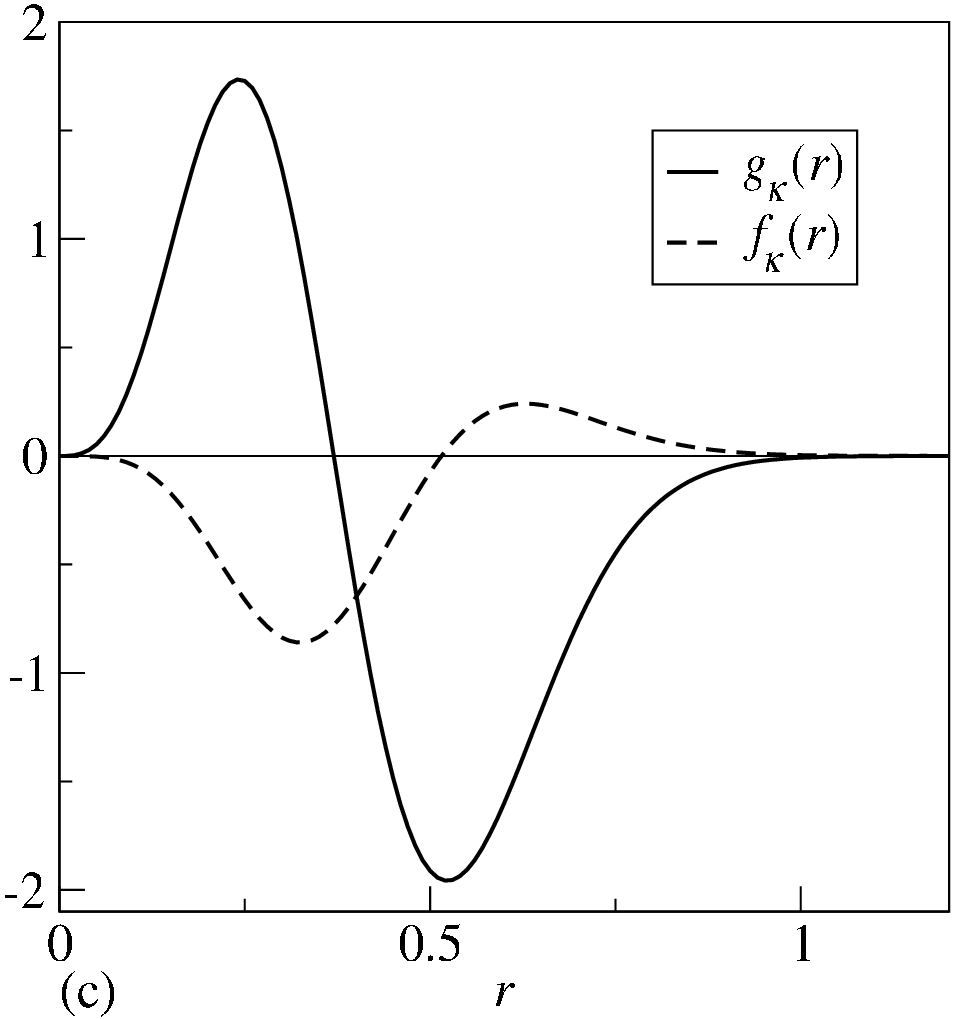}
\end{center}}
}
\vspace*{-0.2cm}
\caption{Radial wave functions for $\Delta=0$
of the
states (a) $2s_{1/2}$, (b) $1d_{3/2}$, and (c) $1d_{5/2}$ with $\omega_{1}=2$, $%
\omega_{2}=1$, and $m=10$.\hfill} \label{Fig:wfDel}
\end{figure}

Since the generalized Laguerre polynomials of degree $n$ have $n$
distinct zeros \cite{abramowitz} we may conclude that $g_{\kappa}$
has $n$
nodes, and the expression for $f_{\kappa}$ suggests that it has $n$ nodes for $%
\kappa<0$ and $n+1$ for $\kappa>0$,
\begin{eqnarray}
n_{f}=\left\{
\begin{array}{ll}
n_{g}\,,   &\;\ \ \ \kappa<0\\[0.2cm]
n_{g}+1\,,   &\;\ \ \ \kappa>0.
\end{array}
\right.\label{Eq:nodes_Delta0}
\end{eqnarray}
This is indeed verified by the radial functions for the states $2s_{1/2}$, $%
1d_{3/2}$, and $1d_{5/2}$ plotted in Fig.~\ref{Fig:wfDel}.

Next, let us consider the other case
\begin{equation}
\Sigma(r) = 0 \, , \ \ \ \Delta(r) = \frac{1}{2} m\, \omega_{1}^2\, r^{2}\,
, \ \ \ U(r) = m\, \omega_{2}\, r\, ,  \label{poteny}
\end{equation}
where $\omega_2$ is again the frequency related to the tensor
potential $U$, and $\omega_1$ is the frequency now related to the
``difference" potential $\Delta$. We now start from the equation
for the lower component, Eq.~(\ref{Eq:D2ordfOHGeral}), and
introduce the new variable and parameters defined by
\begin{eqnarray}
\tilde y &=&\sqrt{\frac{m(\mathcal{E}-m)}{2}}\,\tilde\Omega\,r^{2}=\tilde{a}%
^{2}\,r^{2}\label{tilde_y}\,, \\[0.5cm]
\tilde\Omega&=&\sqrt{\frac{2m}{\mathcal{E}-m}\omega_{2}^2+\omega_{1}^2}\label{tilde_Omega}\,, \\%
[0.5cm]
\tilde\lambda&=&\frac{(2\kappa+1)m\omega_{2}-(\mathcal{E}^2-m^2)}{\tilde{a}%
^{2}}\,.
\end{eqnarray}

The resulting equation for $f_{\kappa }$ in these variables is
formally the same as Eq.~(\ref{Eq:D2ordg(y)}). Following the same
steps as before, to guarantee that $f_{\kappa }(r)$ vanishes when
$r\to \infty $, we must have
\begin{equation}
\tilde{n}=-\frac{1}{2}\biggl(\tilde{l}+\frac{3}{2}+\frac{\tilde{\lambda}}{2}%
\biggr) \ ,  \label{cond_lambda_tilde}
\end{equation}
where $\tilde n$ is an integer greater than or equal to zero and
$\tilde{l}$ is given by Eq.~(\ref{l_tilde}). This last quantum
number is called pseudoorbital angular momentum quantum number in
view of its role in the pseudospin symmetry, which will be
discussed in Sec.~\ref{PSymmetry}. Similarly to Eq. (\ref{5})
for $\kappa $, we may define $\tilde{\kappa}$ in terms of $\tilde{l}$ and $%
\tilde{s}=1/2$ (pseudospin quantum number):
\begin{equation}
\tilde{\kappa}=\left\{
\begin{array}{ccc}
-(\tilde{l}+1) & =-\left( j+1/2\right) , & j=\tilde{l}+1/2\;\ \ \ (\tilde{%
\kappa}<0) \\[0.2cm]
\tilde{l} & =+\left( j+1/2\right) , & j=\tilde{l}-1/2\;\ \ \ (\tilde{\kappa}%
>0)
\end{array}
\ \
\begin{array}{l}
\text{\textrm{aligned pseudospin}} \\
\text{\textrm{unaligned pseudospin.}}
\end{array}
\right.   \label{al_unal_pspin}
\end{equation}
In other words, $\tilde{\kappa}=-\kappa $ is a parameter describing the
coupling between the pseudo-orbital angular momentum and the pseudospin.

\begin{figure}[!ht]
\begin{center}
{\large \includegraphics[width=9cm]{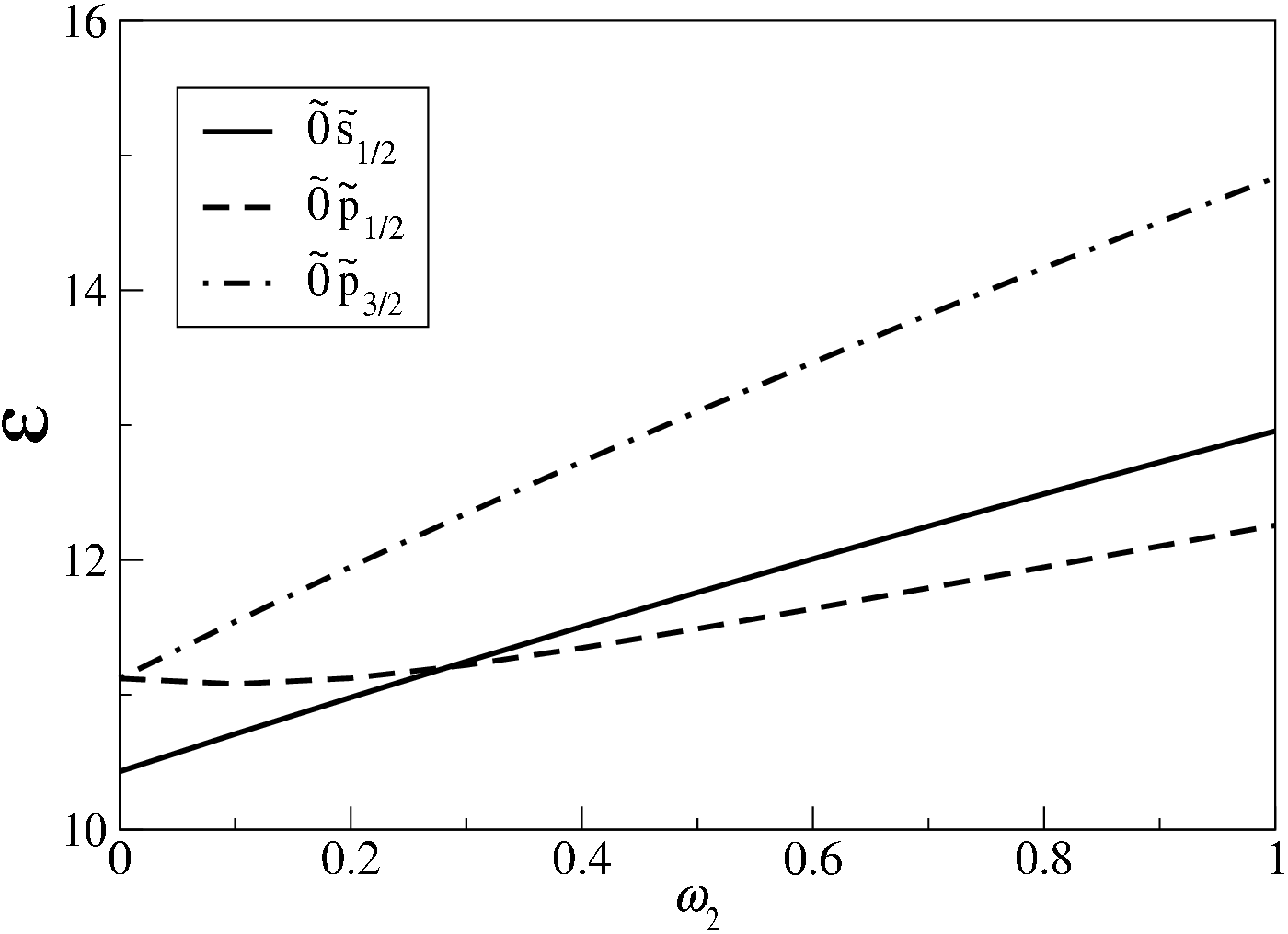}  }
\end{center}
\par
{\large \vspace*{-0.4cm}  } \caption{Energy levels for $\Sigma=0$
with  $\omega_{1}=2$ and $m=10$ as a function of $\omega _{2}$.}
\label{Fig:enSig}
\end{figure}

\begin{figure}[!ht]
\begin{center}
{\large \includegraphics[width=9cm]{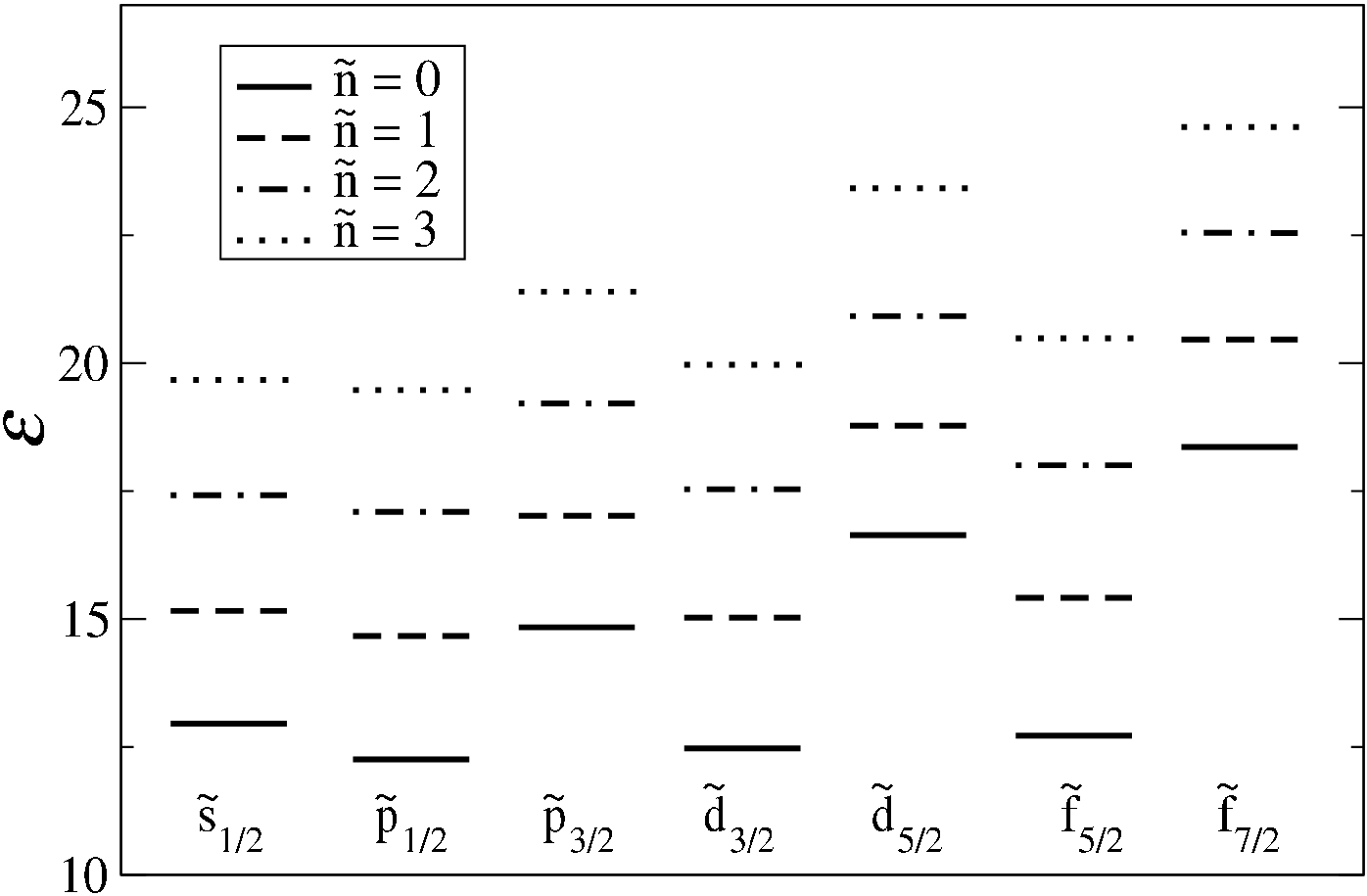}  }
\end{center}
\par
{\large \vspace*{-0.4cm}  } \caption{Energy spectrum for
$\Sigma=0$ with $\omega_{1}=2$, $\omega_{2}=1$, and $m=10$.}
\label{Fig:spcSig}
\end{figure}

From Eq.~(\ref{cond_lambda_tilde}) we get the following equation
for the energy
\begin{equation}
\mathcal{E}^2-m^2+(2\tilde\kappa-1)m\omega_2=\left( 2\tilde{n}+ \tilde{l}+%
\frac{3}{2}\right)\sqrt{2m(\mathcal{E}-m)\omega_1^2+4m^2\omega_2^2} \,.
\label{Eq:AutS0}
\end{equation}
Again, there are at most two real solutions of this equation. For
positive $\omega_2$ and for most cases when $\omega_2<0$, it can
be shown that there is always a positive binding energy solution.
This is the solution we will take in the following examples. Some
energy levels for $\omega_{1}=2$ and $\omega_{2}$ in a range from
$0$ to $1$ are shown in Fig.~\ref{Fig:enSig}. For $\omega_{2}=0$
there is a $(2\tilde{n}+\tilde{l})$ degeneracy of what we might
call a pseudoharmonic oscillator. For a nonvanishing
$\omega_{2}$ this degeneracy is removed as we can also see in
Fig.~\ref{Fig:spcSig}. In these figures we use the spectroscopic
notation $\tilde n\,\tilde l_j\,$, where states with $\tilde
l=0,1,2,\ldots\,$ are denoted by $\tilde s\,,\tilde p\,,\tilde
d\,,\ldots$, respectively. The ground state for $\omega_1=2$ and
$\omega_2=0$ is the state $\tilde 0\tilde s_{1/2}$, but we see
from Fig.~\ref{Fig:enSig} that it is replaced by the state $\tilde
0\tilde p_{1/2}$ as the ground state when $\omega_1=2$ and
$\omega_2\gtrsim 0.3$.

The nonrelativistic limit of the eigenvalue equation in
Eq.~(\ref{Eq:AutS0}) is again reached by decreasing
$\omega _{1}$ and $\omega _{2}$ such that $%
\omega _{1}/m\ll 1$ and $\omega _{2}/m\ll 1$. If we solve the quartic
eigenvalue equation for $E/m$ and expand it in powers of $\omega _{1}/m$
and $\omega _{2}/m$ we get, in lowest order of these expansion parameters,
\begin{equation}
E=\bigg( 2\tilde{n}+\tilde{l}+\frac{3}{2}\bigg)|\omega _{2}|-\bigg(\tilde{%
\kappa}-\frac{1}{2}\bigg)\omega _{2}+\frac{1}{4m}\bigg( 2\tilde{n}+\tilde{l}+%
\frac{3}{2}\bigg) \bigg[2\tilde{n}+\tilde{l}+\frac{3}{2}-\mathrm{sign}%
(\omega _{2})\bigg(\tilde{\kappa}-\frac{1}{2}\bigg)\bigg]\omega _{1}^{2}\ .
\label{Eq:E-non-rel-Sigma_0}
\end{equation}
This equation, valid for $\Sigma =0$ and $\omega_2\not=0$, shows
that the effect of the harmonic-oscillator $\Delta $ potential is
only of second order in a nonrelativistic expansion. This
important result sheds light on the relativistic nature of the
pseudospin symmetry to be discussed later in this paper.

A comparison between Figs.~\ref{Fig:enDel} and \ref{Fig:enSig}
shows that the effect of the tensor potential (seen by the
variation of $\omega_2$), is bigger for the case $\Sigma=0$ than
is for $\Delta=0$. As a consequence, we still see in
Fig.~\ref{Fig:spcSig} a quasidegeneracy for some levels, in
particular, those with the same value of $\tilde n$ and that have
$\tilde l=\tilde\kappa$ (aligned spin). This can be explained by
the energy dependence on $\omega_{1}^2/m$ in
Eq.~(\ref{Eq:E-non-rel-Sigma_0}) (for small values of $\omega
_{1}/m$ and $\omega _{2}/m$), a number smaller than $\omega_{2}$,
meaning that in this case we are near the usual Dirac oscillator
($\omega_1=0$ and $\omega_{2}\neq 0$), where the states with
aligned spin ($j=l+1/2$) and the same $n$ are all degenerate, as
we will discuss in the following section.

The lower component radial wave function is
\begin{equation}
f_\kappa(r)=B\;\mathrm{exp}\Bigr(-\frac{1}{2}\,\tilde{a}^2\,r^{2}\Bigl)\ (%
\tilde{a}^2\,r^{2})^{(\tilde{l}+1)/2} L^{\tilde{l}+1/2}_{\tilde{n%
}}(\tilde{a}^2\,r^{2}) \, ,  \label{Eq:Frfull}
\end{equation}
where $B$ is again a normalization constant, determined by
Eq.~(\ref{4a}). As already commented, the degree of the
generalized Laguerre polynomial determines its number of zeros, so
that $\tilde n$ gives the number of nodes in $f_\kappa$.

The upper component for the set of potentials in
Eq.~(\ref{poteny}) is obtained
from the first order differential equation in Eq.~(\ref{Eq:D1ordRadlow}), i.e.,%
\begin{equation}
g_\kappa(r)=-\frac{1}{\mathcal{E}-m}\biggr[\frac{d \ }{d r}%
+\frac{\tilde\kappa}{r} -m\omega_2\, r \biggl]f_\kappa(r)  \label{Eq:gfull}
\end{equation}
whose explicit form is given, using again the recursion relations
of the Laguerre polynomials, for $\tilde\kappa<0$, by
\begin{equation}
g_\kappa(r)=-\frac{B\,\tilde{a}}{\mathcal{E}-m}\;\exp \Big(\!-\frac{1}{2}%
\,\tilde a^2\,r^{2}\Big) \big(\tilde a^2\,r^{2}\big)^{(\tilde l+2)/2} %
\bigg[\bigg(1-\frac{m\omega_2}{\tilde a^2}\bigg) L^{\tilde l+1/2}%
_{\tilde n} (\tilde a^2\,r^{2}) -2\,L^{\tilde l+3/2}_{\tilde n}
(\tilde a^2\,r^{2})\bigg] ,
\end{equation}
and, for $\tilde\kappa>0$,
\begin{eqnarray}
g_\kappa(r)&=&-\frac{B\,\tilde a}{\mathcal{E}-m}\;\exp \bigg(\!-\frac{1}{2}%
\,\tilde a^2\,r^{2}\bigg)\ \big(\tilde a^2\,r^{2}\big)^{\tilde l/2}%
  \nonumber \\
&\times&\bigg[ \bigg(\tilde n+\tilde l+\frac12\bigg)\bigg(1-\frac{m\omega_2}{%
\tilde a^2}\bigg)L^{\tilde l-1/2}_{\tilde n} (\tilde a^2\,r^{2})
+(\tilde n+1)\bigg(1+\frac{m\omega_2}{\tilde a^2}\bigg)\,L^{\tilde l-1/2}%
_{\tilde n+1} (\tilde a^2\,r^{2})\bigg]\ .
\end{eqnarray}

These functions are analogous to the $g_{\kappa}$ and $f_{\kappa}$
functions of the previous case, so from the above discussion we
can conclude that the number of nodes of the radial functions for
$\Sigma=0$ is such that (remember that $\tilde\kappa=-\kappa$)
\begin{eqnarray}
n_{f}=\left\{
\begin{array}{ll}
n_{g}-1\,, &\;\ \ \ \kappa<0\\[0.2cm]
n_{g}\,,   &\;\ \ \ \kappa>0.
\end{array}
\right.\label{Eq:nodes_Sigma0}
\end{eqnarray}

\begin{figure}[!ht]
{
\parbox[!ht]{5.5cm}{
\begin{center}
\includegraphics[width=5.5cm,height=5.5cm]{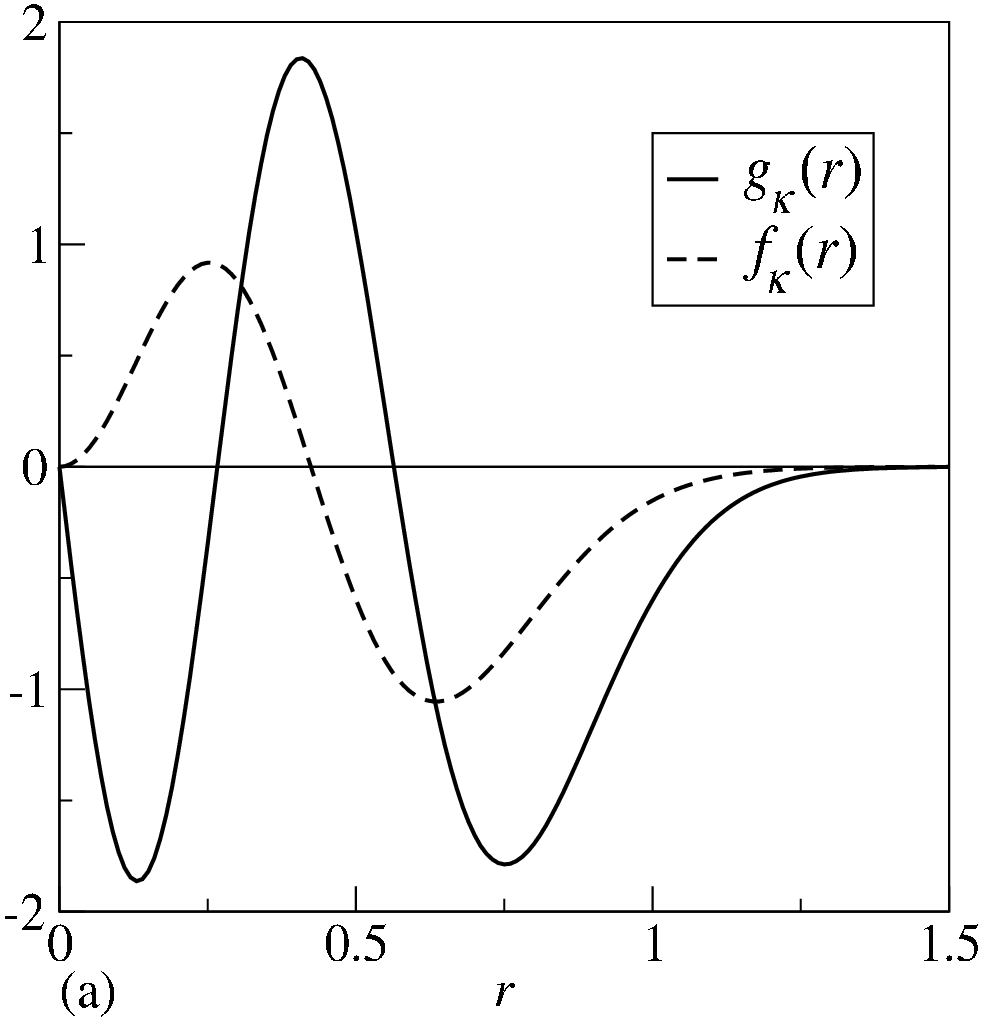}
\end{center}
}\hfill
\parbox[!ht]{5.5cm}{
\begin{center}
\includegraphics[width=5.5cm,height=5.5cm]{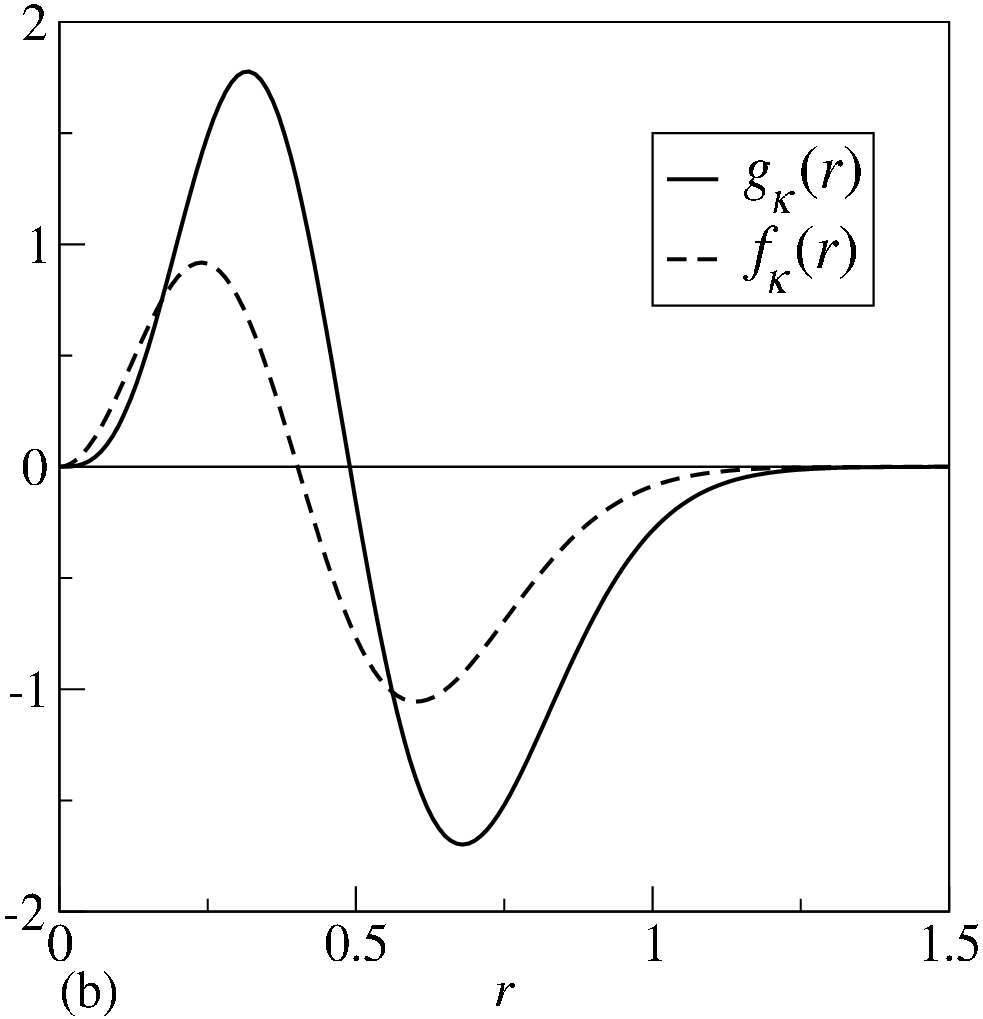}
\end{center}
}\hfill
\parbox[!ht]{5.5cm}{
\begin{center}
\includegraphics[width=5.5cm,height=5.5cm]{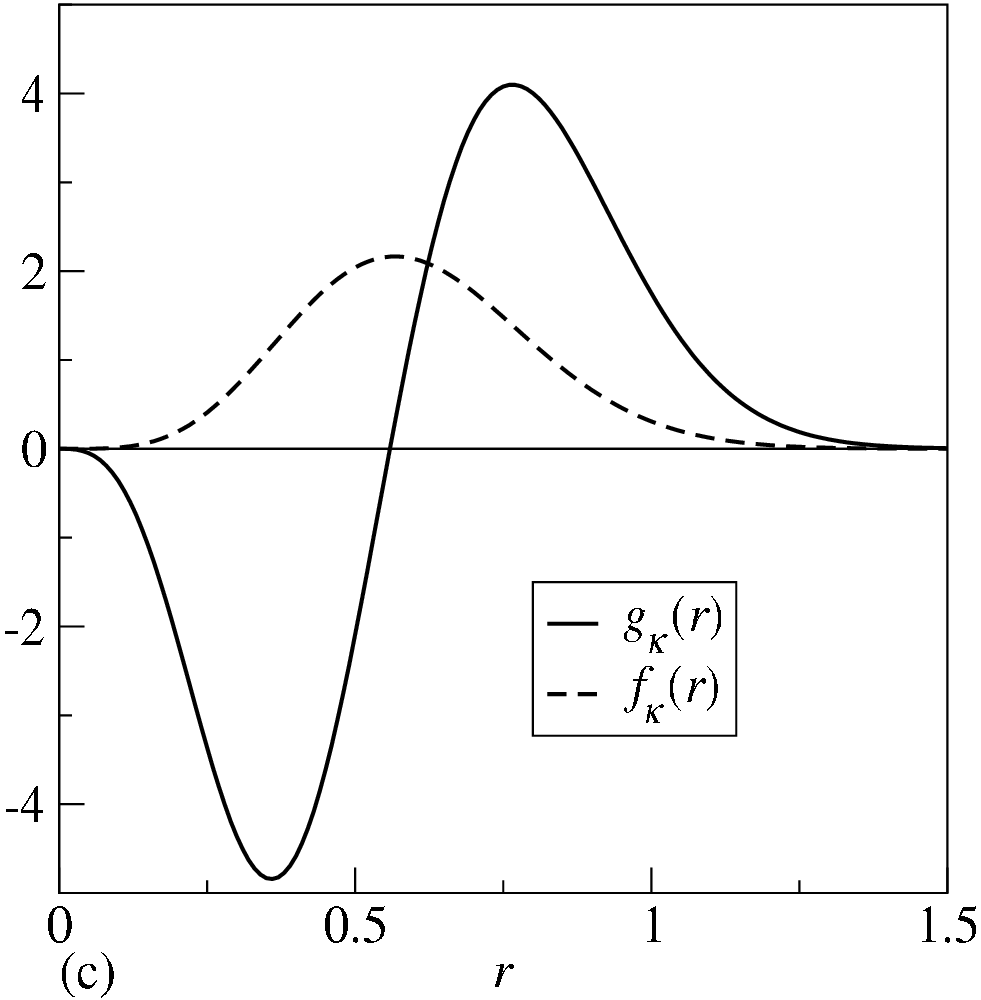}
\end{center}
}  } \vspace*{-.2cm} \caption{ Radial wave functions for
$\Sigma=0$ of the states (a) $\tilde 1\tilde p_{1/2}$
($2s_{1/2}$), (b) $\tilde 1\tilde p_{3/2}$ ($1d_{3/2}$), and (c)
$\tilde 0\tilde f_{5/2}$ ($1d_{5/2}$) with $\omega_{1}=2$, $
\omega_{2}=1$, and $m=10$.\hfill\ } \label{Fig:wfSig}
\end{figure}

This is illustrated in the plots of $g_\kappa$ and $f_\kappa$ in Fig.~%
\ref{Fig:wfSig}. Adopting the definition of $\tilde n$ as the
number of nodes of the lower component when $\omega_2=0$ and
$\Sigma=0$, we see that we can convert from the notation $n\,l_j$
to $\tilde n\,\tilde l_j$ for the same state by setting
$n\to\tilde n=n,\,l\to\tilde l=l-1$ when $\kappa>0$ and
$n\to\tilde n=n-1,\,l\to\tilde l=l+1$ when $\kappa<0$.

We have noticed, however, that the radial function node structure
depends on the ratio $\omega_1/\omega_2$. If $\omega_2$ becomes
significantly larger than $\omega_1$, in the case of $\Delta=0$
the structure of the nodes is the one of the normal Dirac
oscillator, the same as in Eq.~(\ref{Eq:nodes_Sigma0}), as it will
be discussed in the following section.


\section{\protect Particular cases}

Now we will present some results for the three particular cases
$\omega_1 = 0$, $\omega_2 = 0$ ($\Delta=0$ {\text and}
$\Sigma=0$).

\subsection{\protect Case $\omega_1 = 0$}

\label{SubSec:OscDirac}

As mentioned before, the expression ``Dirac oscillator" applies to
this case when $\omega_2>0$ \cite{Mosh_S}. Here we also consider
the possibility that $\omega_2$ be negative. From Eq.~(\ref{ggg})
when $\omega_1 = 0$, it is straightforward to conclude that the
equation for the upper radial wave function of the Dirac
oscillator is a Schr\"odinger-like equation with a 
harmonic-oscillator potential and a spin-orbit coupling term.

\begin{figure}[!ht]
\begin{center}
\includegraphics[width=9cm]{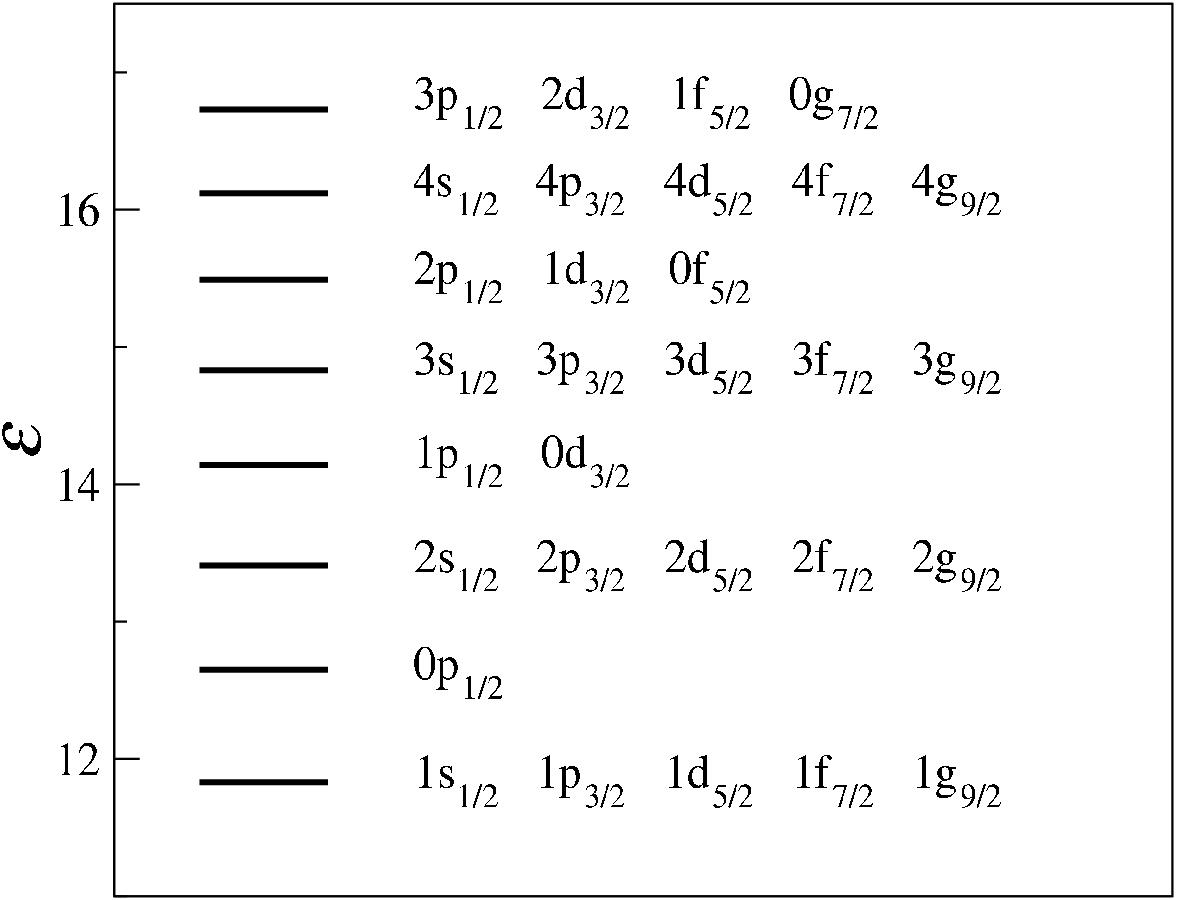}
\end{center}\vspace*{-0.4cm}
\caption{Energy spectra for the normal Dirac oscillator
($\omega_{1}=0$) with $\omega_{2}=1$ and $m=10$.\hfill }
\label{Fig:spcReSig2}
\end{figure}

The eigenenergies are readily obtained taking the limit $%
\omega_{1}\rightarrow\, 0$ in Eq.~(\ref{Eq:AutD0}), which yields
\begin{equation}
\mathcal{E}^2=m^2+m|\omega_{2}| \left[2\left( N+ \frac{3}{2}\right)+\mathrm{%
sgn}(\omega_2)(2\kappa-1) \right] \, ,  \label{eneros1}
\end{equation}
where the quantum number $N=2n+l$ was introduced. There are, in
general (when the right-hand side of the equation is positive) two
symmetric real solutions for this equation. Here we take the
positive energy solution. Note that Eq.~(\ref {eneros1}) results
from the equation for the upper component, $g_\kappa$.
We can get the nonrelativistic energy by setting $\omega_1 = 0 $ in Eq.~(%
\ref{Eq:E-non-rel-Delta_0}). Therefore, in this limit we obtain
for the energy $E$ the quantum mechanical result
$|\omega_2|(N+3/2)$ plus the spin-orbit contribution
$\frac{1}{2}(2\kappa-1)\omega_{2}$. This term removes the
nonrelativistic degeneracy related to levels with the same $N$ in
such a way that changing the sign of $\omega_2$ leads to a order
reversal of the levels
with the same $(n,\,l\,)$ and $j=l\pm 1/2$ quantum numbers. For $%
\omega_2>0$, depending on whether the quantum number $\kappa$ is positive or
negative, the energy is given by [see Eq.~(\ref{5})]
\begin{eqnarray}  \label{Eq:EnrOHRG}
\mathcal{E}^{2}=m^{2}+4\,m\omega_{2}\left(n+l+\frac{1}{2}\right),&\qquad%
\kappa>0, \\[0.5cm]
\mathcal{E}^{2}=m^{2}+4\,m\omega_{2} n,&\qquad\kappa<0.
\label{Eq:EnrOHRGkneg}
\end{eqnarray}
Thus, for a given quantum number $n$, all states with aligned spin
$j=l+1/2 \ (\kappa<0)$ and \textit{arbitrary} orbital angular
momentum are degenerate. This degeneracy, as well as the $(n+l)$
degeneracy for $\kappa>0$, is illustrated in
Fig.~\ref{Fig:spcReSig2}, in which the spectrum for $\omega_{2}=1$
is presented.
On the other hand, the eigenenergies of the Dirac oscillator
Hamiltonian may also be determined from Eq.~(\ref{Eq:AutS0}),
which was obtained from the second order equation for the lower
component $f_{\kappa}$, very similar to Eq.~(\ref{ggg}). Again taking the limit $%
\omega_{1}\rightarrow\, 0$ in Eq.~(\ref{Eq:AutS0}) one obtains
\begin{equation}
\mathcal{E}^2=m^2+m|\omega_{2}|\left[2\left(\tilde{N}+\frac{3}{2}\right)+%
\mathrm{sgn}(\omega_2)(2\kappa+1)\right],  \label{42}
\end{equation}
where $\tilde{N}=2\tilde{n}+\tilde{l}$ is a new quantum number. The term $%
2\kappa+1$ comes from the spin-orbit interaction for the lower component.
Depending on whether the quantum number $\kappa$ is positive or negative,
the energy, for $\omega_2>0$, is given by
\begin{eqnarray}
\mathcal{E}^{2}=m^{2}+4m\omega_{2}\left( \tilde{n}+\tilde{l}+\frac{3}{2}%
\right),&\qquad\kappa>0,  \label{Eq:EnrOHRF} \\[0.5cm]
\mathcal{E}^{2}=m^{2}+4 m \omega_{2}\, (\tilde{n}+1),&\qquad\kappa<0.
\label{Eq:EnrOHRFkneg}
\end{eqnarray}

Of course, the eingenvalue equations in Eqs.~(\ref{Eq:EnrOHRF}) and
(\ref{Eq:EnrOHRFkneg}) should lead to the same energy values as
Eqs.~(\ref{Eq:EnrOHRG}) and (\ref{Eq:EnrOHRGkneg}).
Remembering that, for $\kappa<0$, one has $\tilde{l} =l+1$, and,
for $\kappa>0$, $\tilde{l}=l-1$, from the comparison of those two
sets of equations one concludes that
\begin{eqnarray}
n_{f}=\left\{
\begin{array}{ll}
n_{g}-1\,, &\;\ \ \ \kappa<0\\[0.2cm]
n_{g}\,,   &\;\ \ \ \kappa>0,\label{Eq:nOHR}
\end{array}
\right.
\end{eqnarray}
since the quantum numbers $n$ and $\tilde n$ are equal to the
number of nodes of the upper and lower radial functions, denoted
by $n_{g}$ and $n_{f}$, respectively. Notice that $n_g\geq 1$ for
$\kappa<0$. This can also be inferred from the analytical form of
the radial upper and lower functions for the Dirac oscillator.
These components are readily obtained from Eqs.~(\ref{Eq:Grfull}),
(\ref{Eq:Gen_f_kappa_neg}), and (\ref{Eq:Gen_f_kappa_pos}),
noticing that, when $\omega_1=0$, the square of the parameter $a$,
defined in Eq.~(\ref{xx3}), is equal to $m\omega_2$
($\omega_2>0$), giving
\begin{equation}
\label{Eq:G_OscD}
g_\kappa(r)=A\;\exp\Bigr(\!-\frac{1}{2}\,a^2\,r^{2}\Bigl)\ (a^2\,r^{2})^{(l+1)/2}
\ L^{l+1/2}_{n} (a^2\,r^{2})
\end{equation}
for the upper component, and
\begin{eqnarray}
f_\kappa(r)&=&-\frac{2A\, a}{\mathcal{E}+m}\;\exp \Big(\!-\frac{1}{2}%
\,a^2\,r^{2}\Big) \big(a^2\,r^{2}\big)^{(l+2)/2} \,L^{l+3/2}%
_{n-1} (a^2\,r^{2})\qquad(\kappa<0)
\label{Eq:F_OscD_kappa_neg}\\[0.5cm]
f_\kappa(r)&=&\frac{2A\, a}{\mathcal{E}+m}\;\exp \Big(\!-\frac{1}{2}%
\,a^2\,r^{2}\Big)\ \big(a^2\,r^{2}\big)^{l/2} \bigg(n+l+\frac12\bigg)%
\,L^{l-1/2}_{n} (a^2\,r^{2}) \qquad(\kappa>0)  \label{Eq:F_OscD_kappa_pos}
\end{eqnarray}
for the lower component [we have used the relation
$L_{n}^{\alpha-1}(x) - L_{n}^{\alpha}(x)=-L_{n-1}^{\alpha}(x)$
\cite{abramowitz}].

As we have seen before, the degree of these Laguerre polynomials
is identified with the principal quantum number of the upper and
lower radial functions, and thus with their number of nodes, so
that the relations in Eq.~(\ref{Eq:nOHR}) follow immediately.

\begin{figure}[!ht]
\begin{center}
{\large
\includegraphics[width=8cm]{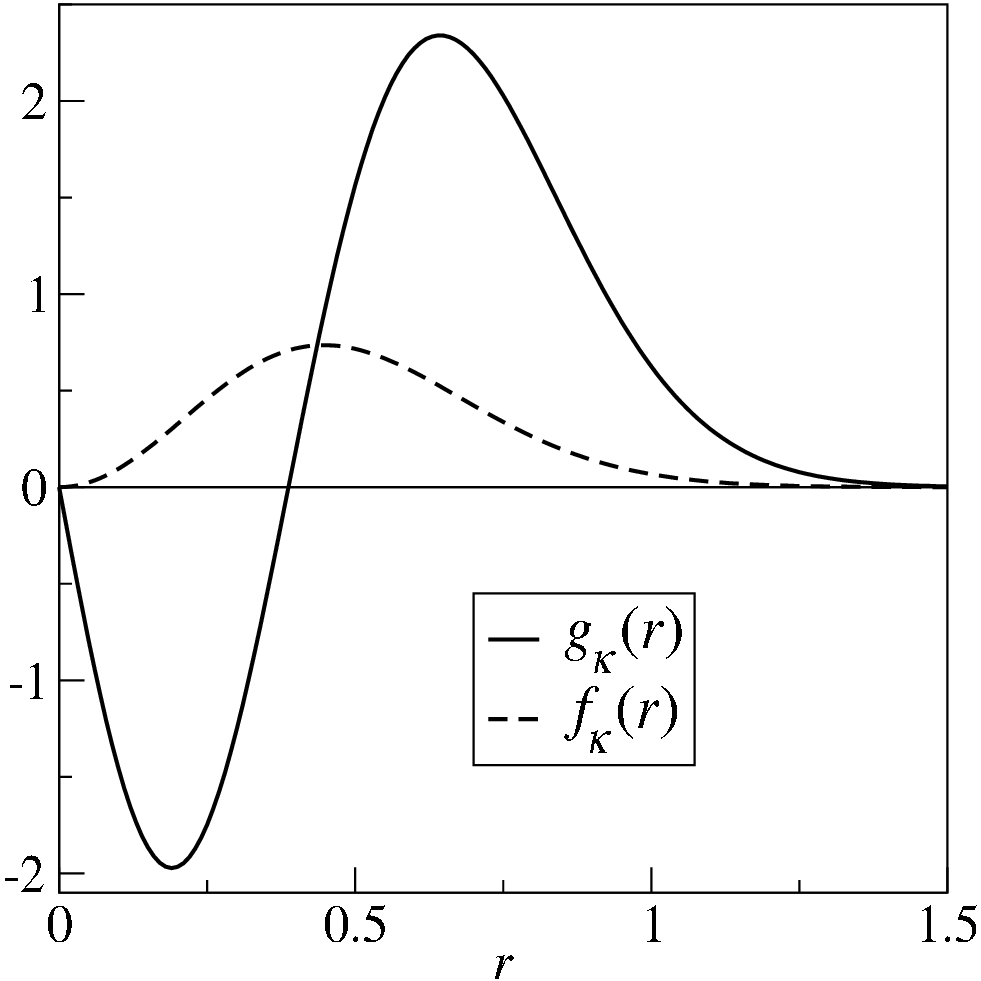}  }
\end{center}
\par
{\large \vspace*{-0.4cm}  } \caption{Ground state radial wave
functions for the normal Dirac oscillator with the same parameters
as in Fig.~\ref{Fig:spcReSig2}.\hfill\ } \label{Fig:n0s12}
\end{figure}

 In this case ($\omega_{2}>0$) Eq.~(\ref{Eq:nOHR}) implies
a peculiar node structure for the ground state $s_{1/2}$ ($\kappa
= -1$). This state has a node ($n_{g}=1$) in the upper component,
as we can see in Fig.~\ref{Fig:n0s12}, whereas the
nonrelativistic ground state wave function is nodeless.

 If $\omega_2<0$, the infinite degeneracy referred to
above occurs for $\kappa>0$, i.e., for nonaligned spins.
In addition, the relation between the values $n_{g}$ and $n_{f}$
is modified: $n_{f}=n_{g}+1$ for $\kappa>0$ and $n_{f}=n_{g}$ for
$\kappa<0$. It is worth noticing that the structure of the radial
nodes for $\omega_2$ positive (negative) is exactly the same for
the case $\Sigma=0$ ($\Delta=0$) discussed in the preceding
section. This same node structure could again be inferred from the
form of the radial wave functions when $\omega_2<0$, which can be
obtained once more from Eqs.~(\ref{Eq:Grfull}), (\ref
{Eq:Gen_f_kappa_neg}), and (\ref{Eq:Gen_f_kappa_pos}), setting $%
a^2=-m\omega_2 $. We note that now the ground state $s_{1/2}$ has
zero nodes in the upper component, and therefore, in the
nonrelativistic limit, when the lower component disappears, the
wave function has the same node structure as the nonrelativistic
harmonic-oscillator ground state wave function. Furthermore, from
Eq.~(\ref{42}) we see that, for $\omega_2<0$, the sign of the
spin-orbit interaction is inverted, such that states with
nonaligned spin ($\kappa>0$) have now lower energy than states
with the same $l$ and aligned spin. This is the situation in
atomic physics, where, for example, the states $p_{3/2}$ have
higher energy than the states $p_{1/2}$. The opposite happens in
nuclear physics, where states with aligned spin are lower in
energy than non-aligned spin states, which in the present case
happens when $\omega_2>0$.
\begin{figure}[!ht]
{
\parbox[!ht]{5.5cm}{
\begin{center}
\includegraphics[width=5.5cm,height=5.5cm]{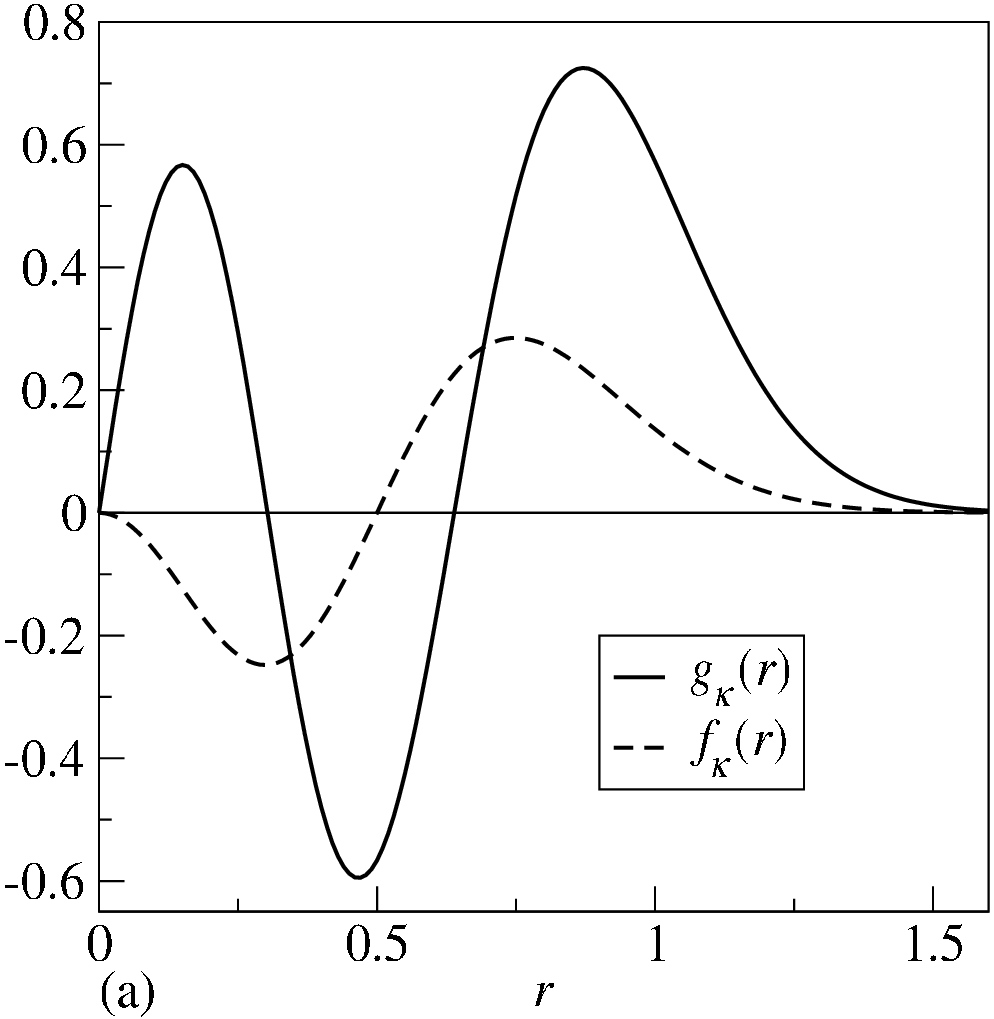}
\end{center}
}\hfill
\parbox[!ht]{5.5cm}{
\begin{center}
\includegraphics[width=5.5cm,height=5.5cm]{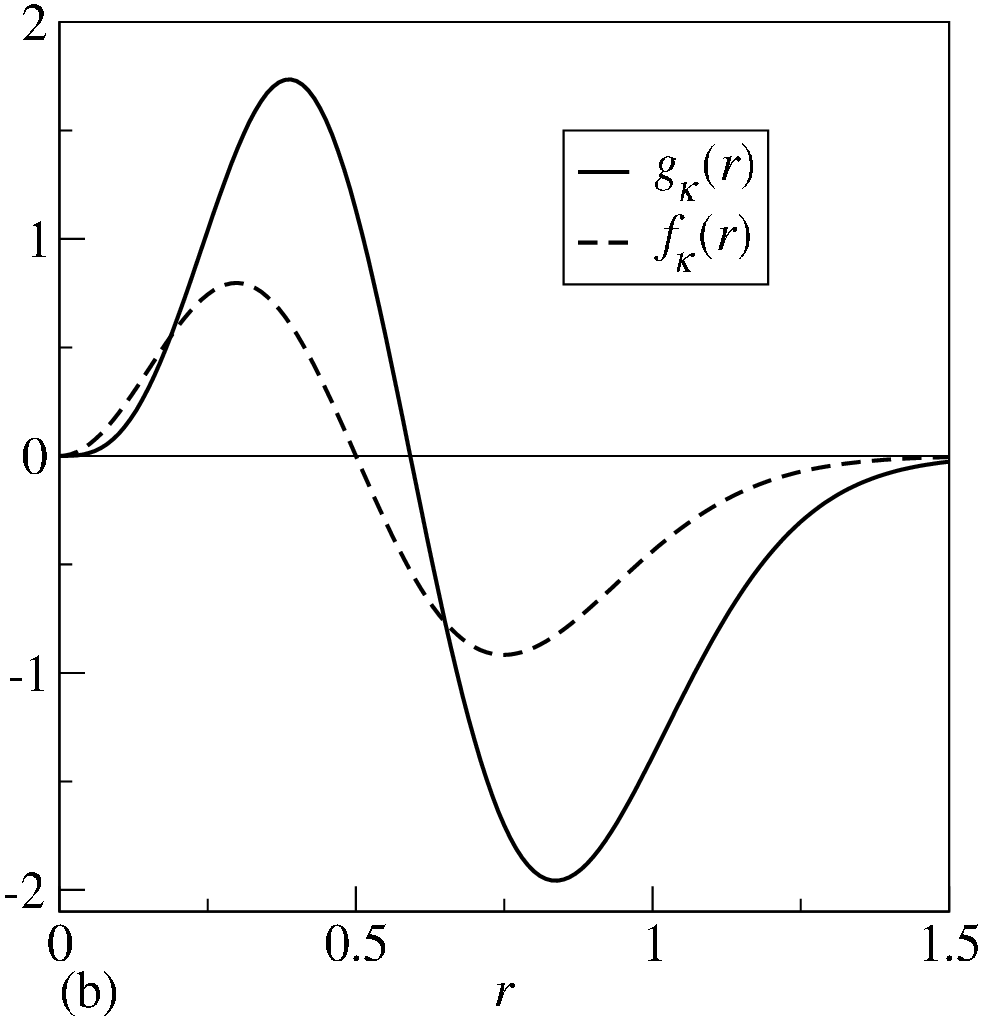}
\end{center} }\hfill
\parbox[!ht]{5.5cm}{
\begin{center}
\includegraphics[width=5.5cm,height=5.5cm]{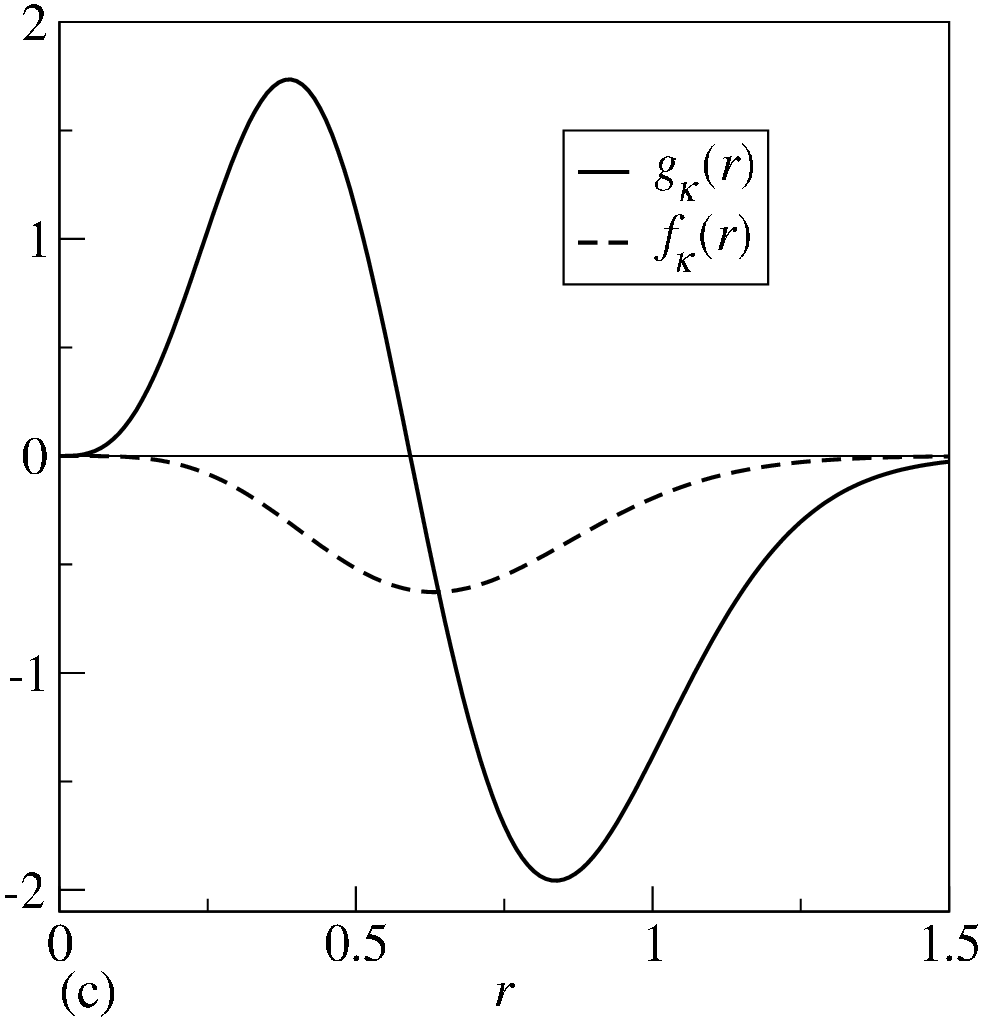}
\end{center}
}  } \vspace*{-.2cm}
 \caption{Radial wave functions for the normal
Dirac oscillator of the states (a) $2s_{1/2}$, (b) $1d_{3/2}$, and
(c) $1d_{5/2}$ with the same parameters as in
Fig.~\ref{Fig:spcReSig2}.\hfill\ } \label{ReDel}
\end{figure}

 We present in Fig.~\ref{ReDel} the radial wave functions for the
states $2s_{1/2}$, $1d_{3/2}$, and $1d_{5/2}$. From these plots we
see that the upper components for the spin-orbit partners (same
$n,l$) are equal [see Fig.~\ref{ReDel}(b) and Fig.~\ref{ReDel}(c)],
since both are given by the same radial
wave function, Eq.~(\ref{Eq:G_OscD}). For pairs with the same $%
\tilde{l}$ (pseudospin partners) the lower components are the same, up to a
constant factor. This can be seen from Eqs.~(\ref{Eq:F_OscD_kappa_pos}) and (\ref
{Eq:F_OscD_kappa_neg}), since the pseudospin partners have quantum numbers $%
(n\,,l)$ ($\kappa <0$) and $(n-1\,,l+2)$ ($\kappa >0$). This
particular behavior could be related to a particular symmetry of
the Dirac Hamiltonian in Eq.~(\ref{2}) with $\omega _{1}=0$, as is
the case with the pseudospin symmetry, to be discussed in
Sec.~\ref{PSymmetry}.


\subsection{\protect Case $\omega_2 = 0$ and $\Delta=0$}

\label{SubSec:omega2_0_D_0}

In this subsection we set to zero both $\Delta$ and $U$
potentials. Then, Eq.~(\ref{ggg}) with $\omega_2 = 0$ leads to
the following second order differential equation for the upper
component [note that $\kappa(\kappa+1)=l(l+1)$,
irrespective of the sign of $\kappa$]
\begin{equation}
\biggl[\frac{\mathrm{d} ^2\ }{\mathrm{d} \,r^2}-\frac{l(l+1)}{r^2}-\frac{m(%
\mathcal{E}+m)}{2}
\omega_{1}^{2}r^{2}-(m^{2}-\mathcal{E}^2)\biggr]g_\kappa(r)=0.
\label{Eq:D2ordgOHr2}
\end{equation}

\begin{figure}[!ht]
{\parbox[!ht]{5.5cm}{
\begin{center}
\includegraphics[width=5.5cm,height=5.5cm]{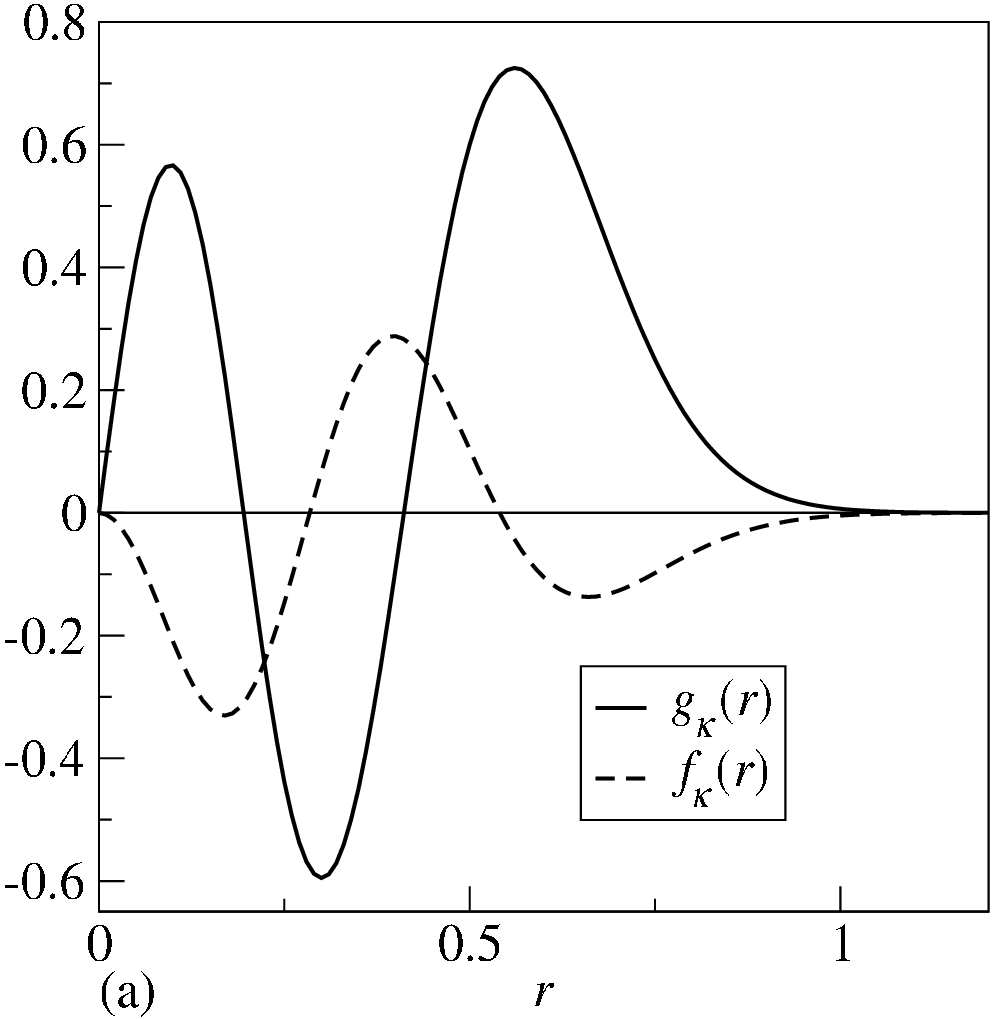}
\end{center}
}\hfill
\parbox[!ht]{5.5cm}{
\begin{center}
\includegraphics[width=5.5cm,height=5.5cm]{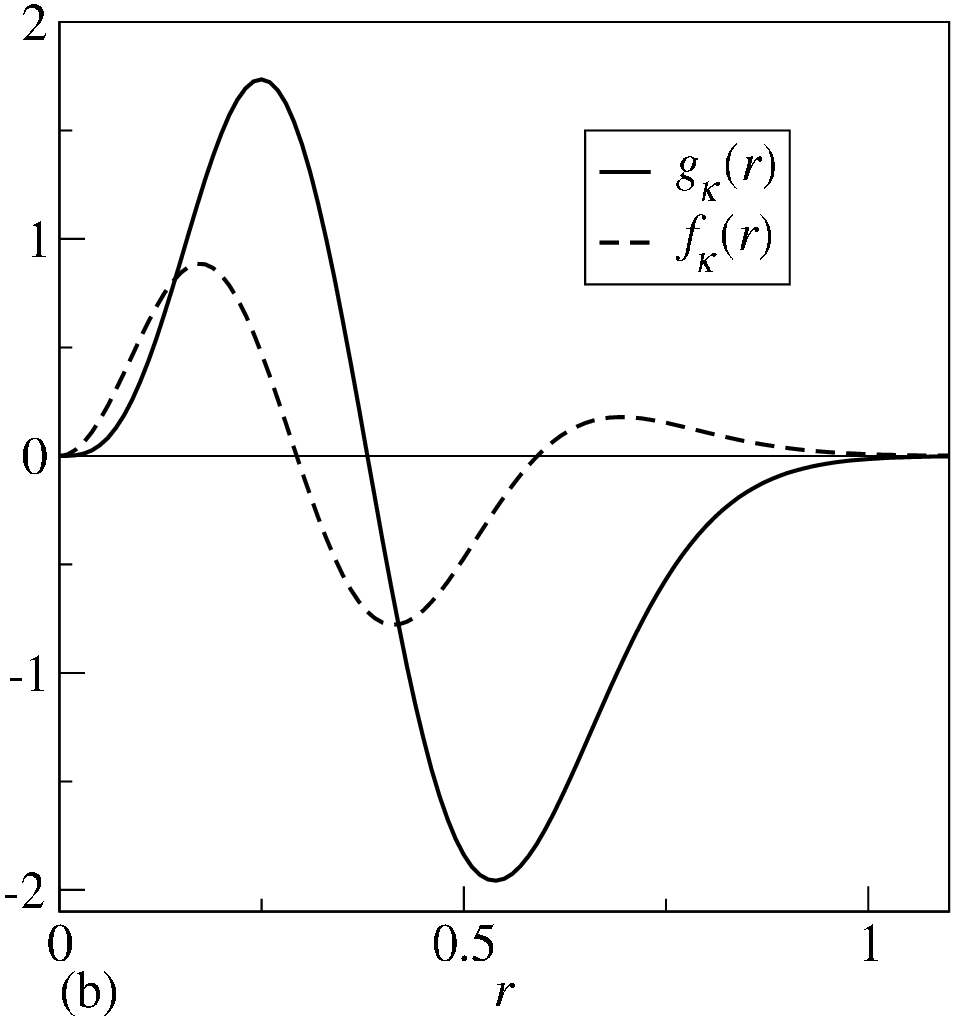}
\end{center}}\hfill
\parbox[!ht]{5.5cm}{
\begin{center}
\includegraphics[width=5.5cm,height=5.5cm]{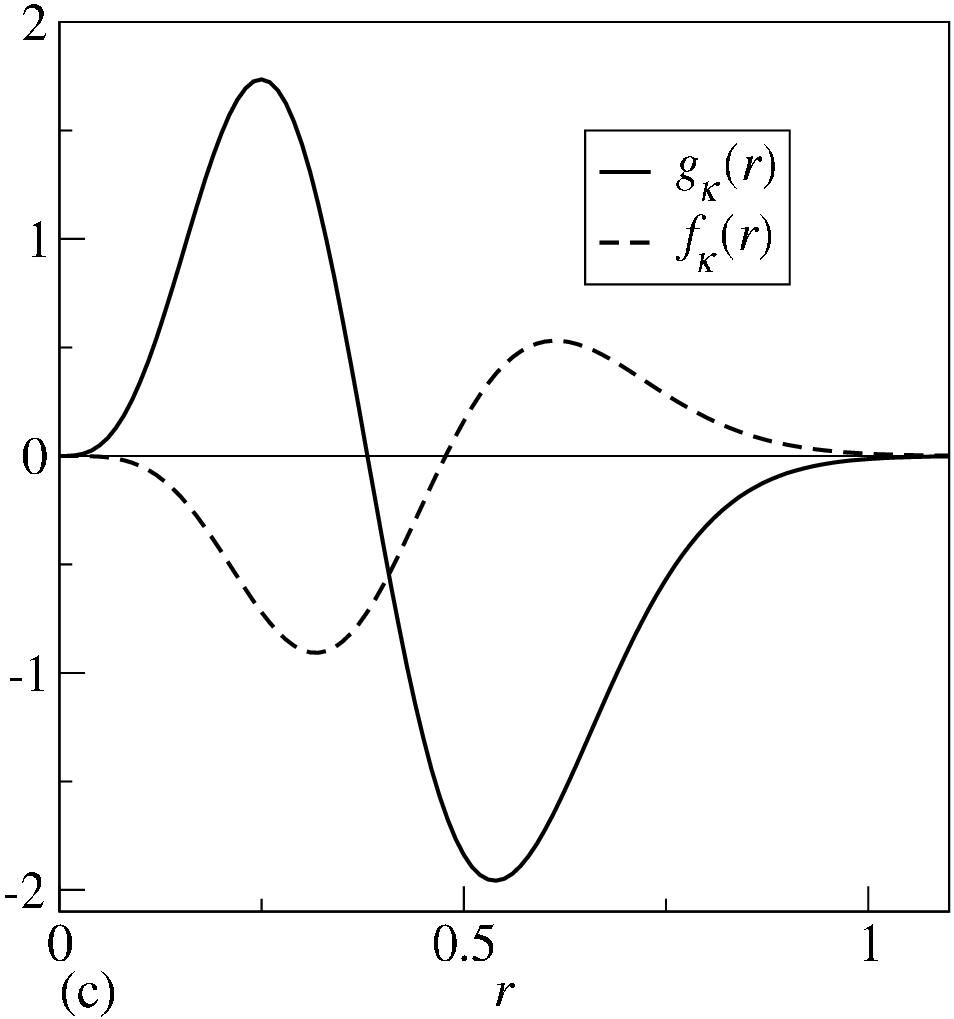}
\end{center}
}  } \vspace*{-0.2cm}
\caption{Radial wave functions for the potential $\Sigma(r)=\frac{1}{2}%
m\omega_{1}^{2}r^2$, with $U=\Delta=0$ for the states $2s_{1/2}$,
$1d_{3/2}$, and $1d_{5/2}$, respectively, with $\omega_{1}=2$ and
$m=10$.\hfill\ } \label{Fig:PoDel}
\end{figure}

The solutions of this equation are given by Eq.~(\ref{Eq:Grfull}),
where $a$ is given by  Eqs.~(\ref{xx1})--(\ref{xx3}) in which
$\omega_{2}$ is set to zero. The upper and lower radial functions
for two levels corresponding to $\kappa$ with opposite signs are
plotted in Fig.~\ref{Fig:PoDel}. From this figure we see that the
node structure is the same as given in the preceding section when
$\omega_{2}\neq 0$ [Eq.~(\ref{Eq:nodes_Delta0})].
The energy eigenvalues are
readily obtained from Eq.~(\ref{Eq:AutD0}) by just taking the limit $%
\omega_{2}\rightarrow\,0$, yielding
\begin{equation}
(\mathcal{E}-m)\sqrt{\frac{\mathcal{E}+m}{2m}}=\omega_{1}\left (2 n+l+\frac{3%
}{2}\right) \qquad(n=0,1,2,\ldots).  \label{Eq:EnrOHpaper}
\end{equation}
This equation for the energy, valid for $\Delta=0$, shows
explicitly that there is no spin-orbit term and that states with
$j=l\pm1/2$ are degenerate. We can prove that this equation has
only one real solution, and it has to be such that
$\mathcal{E}>m$. To begin with, since the right-hand side is always
positive, it is clear that any real solution must be greater than
$m$, otherwise the left-hand side would be negative. We prove that
there is only one of such solutions by squaring
Eq.~(\ref{Eq:EnrOHpaper}) and using the Descartes'{} rule of
signs. This rule states that the number of positive real roots of
an algebraic equation with real coefficients
$a_{k}x^{k}+\cdots+a_{1}x +a_{0}=0$ is never greater than
the number of changes of signs in the sequence $%
a_{k},\ldots,a_{1},a_{0}$ (not counting the null coefficients)
and, if less, then always by an even number \cite{sal,bar}. 
Since we can write
the square of Eq.~(\ref{Eq:EnrOHpaper}) as
\begin{equation}
x^3+2x^2-a^2=0\ ,
\end{equation}
with $x=(\mathcal{E}-m)/m$ and $a=\sqrt{2}\omega_{1}(2
n+l+3/2)/m$, then, by Descartes'{} rule of signs, only one
solution with $x>0$, i.e., $\mathcal{E}>m$, exists. Since by
squaring Eq.~(\ref{Eq:EnrOHpaper}) we may only introduce a
negative ($x<0$) spurious solution, this solution is the only
positive solution of that equation, and the proof is complete.
Therefore, for given values of $\omega_1$, $n$ and $l$, i.e, only
discrete, positive (greater than $m$) energies are allowed.

\begin{figure}[!ht]
\begin{center}
\includegraphics[width=9cm]{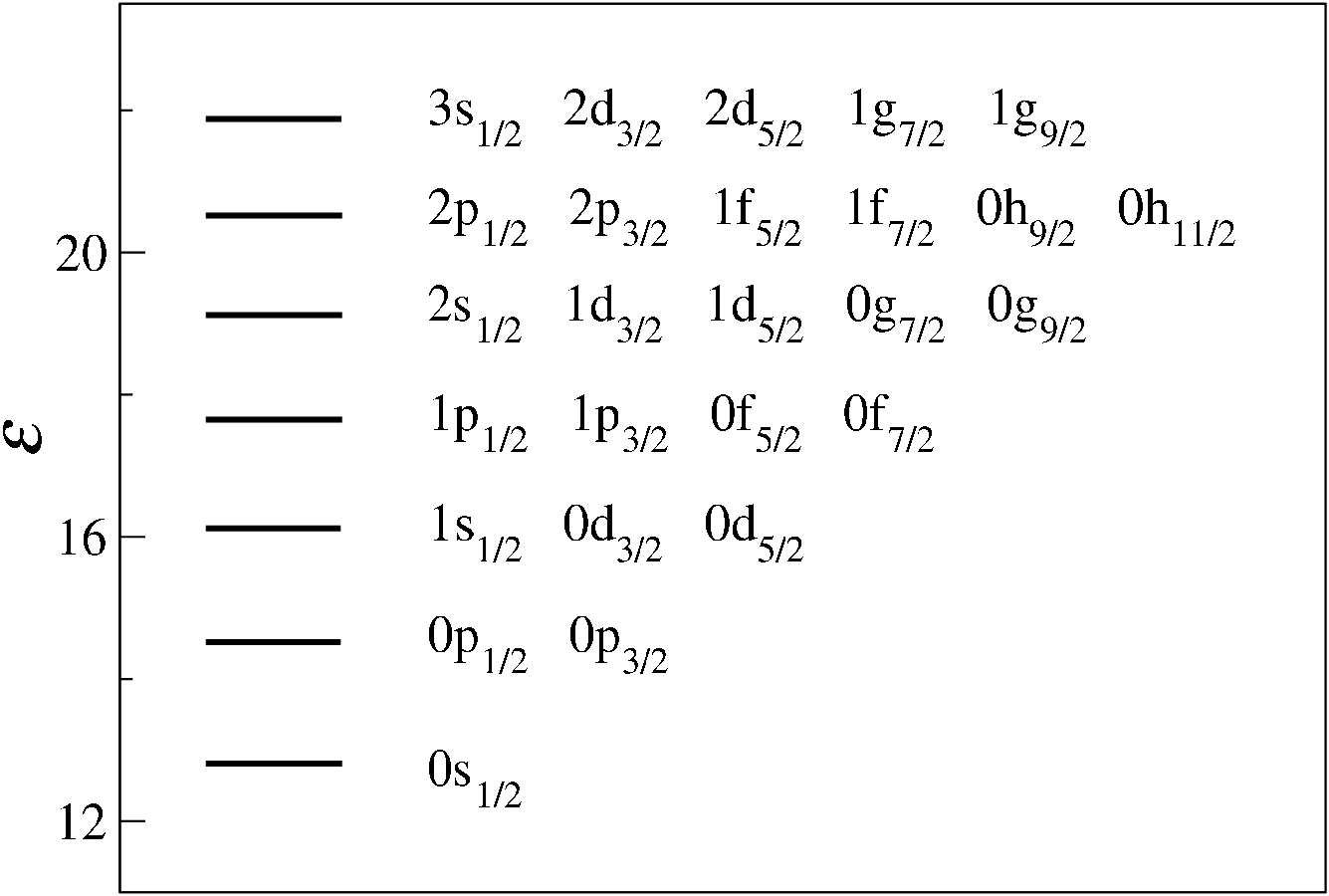}
\end{center}
\par
\vspace*{-0.5cm} \caption{Single particle energies for the case
$\Delta=U=0$ with $\omega_{1}=2$, $\omega_{2}=0$, and $m=10$.} \label{Fig:spcPoDel}
\end{figure}

The nonrelativistic limit is obtained by enforcing $\omega
_{2}\to 0$ in Eq.~(\ref{Eq:E-non-rel-Delta_0}). The right-hand
side becomes the expression for the nonrelativistic energy
eigenvalues of the harmonic-oscillator with no spin-orbit
coupling. Similarly, Eq.~(\ref{Eq:Grfull}) gives the
non-relativistic harmonic oscillator radial wave function, with $%
a^{2}=m\omega_{1}$. Figure \ref{Fig:spcPoDel} shows the single
particle spectrum for $\Delta =0$, $m=10$, and $\omega_{1}=2$. For
sufficiently small $\omega _{1}$ the energy levels are essentially
equidistant, a well-known feature of nonrelativistic harmonic-oscillator 
energy levels, and we do not have spin-orbit
interaction: states with $j=l\pm1/2$ are degenerate. These results
allow us to stress that the $\Delta$ potential can be closely
connected to the intrinsic relativistic content of a theory with
scalar $(S)$ and vector $(V)$ potentials. The nonrelativistic
harmonic oscillator is in fact the limit of the relativistic
theory with $\Delta=V-S=0$ and a potential well $\Sigma=S+V$ when
$\omega_1/m$ becomes very small.

Thus, an important conclusion of our study is that a Dirac
equation with scalar and vector harmonic-oscillator potentials
with the same sign and magnitude ($\Delta=0$), instead of a linear
tensor potential $U$, seems to be the most natural way to
introduce the harmonic oscillator in relativistic quantum
mechanics.

\subsection{\protect Case $\omega_2 = 0$ and $\Sigma=0$}

\label{SubSec:omega2_0_S_0}

Now we consider that both the tensor potential $U$ and $\Sigma$
are zero. The situation is very similar to the preceding
subsection, except that we now find that it is the lower component
that satisfies the following second order differential equation
\begin{equation}
\biggl[\frac{\mathrm{d}^2\ }{\mathrm{d} r^2}-\frac{\tilde{l}(\tilde{l}+1)}{%
r^2}-\frac{m(\mathcal{E}-m)}{2} \omega_{1}^{2}r^{2}-(m^{2}-\mathcal{E}^2)%
\biggr]f_\kappa(r)=0  \label{Eq:D2ordfOFr2}
\end{equation}
[note that $\kappa(\kappa-1)=\tilde\kappa(\tilde\kappa+1)=\tilde l(\tilde%
l+1)$]. The solution of this equation is the function
(\ref{Eq:Frfull}) setting $\omega_{2}=0$ in
Eq.~(\ref{tilde_Omega}).

\begin{figure}[!ht]
\parbox[!ht]{5.5cm}{
\begin{center}
\includegraphics[width=5.5cm,height=5.5cm]{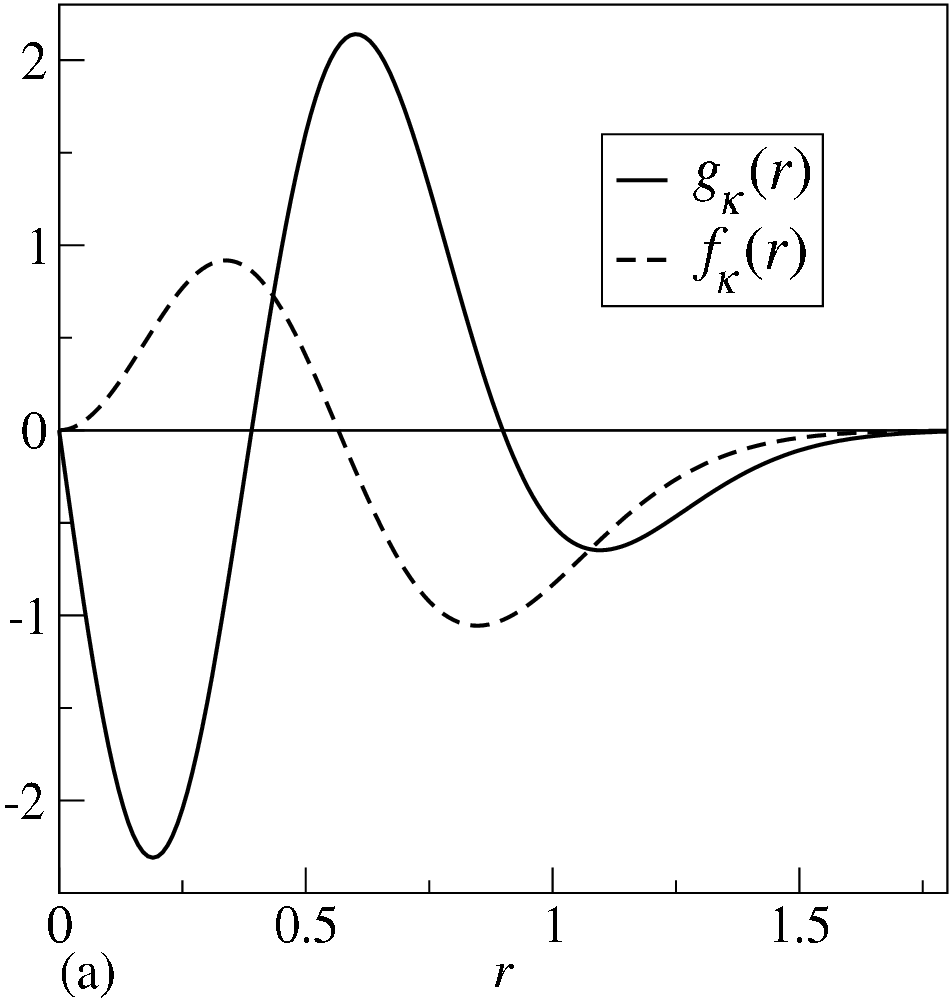}
\end{center}
}\hfill
\parbox[!ht]{5.5cm}{
\begin{center}
\includegraphics[width=5.5cm,height=5.5cm]{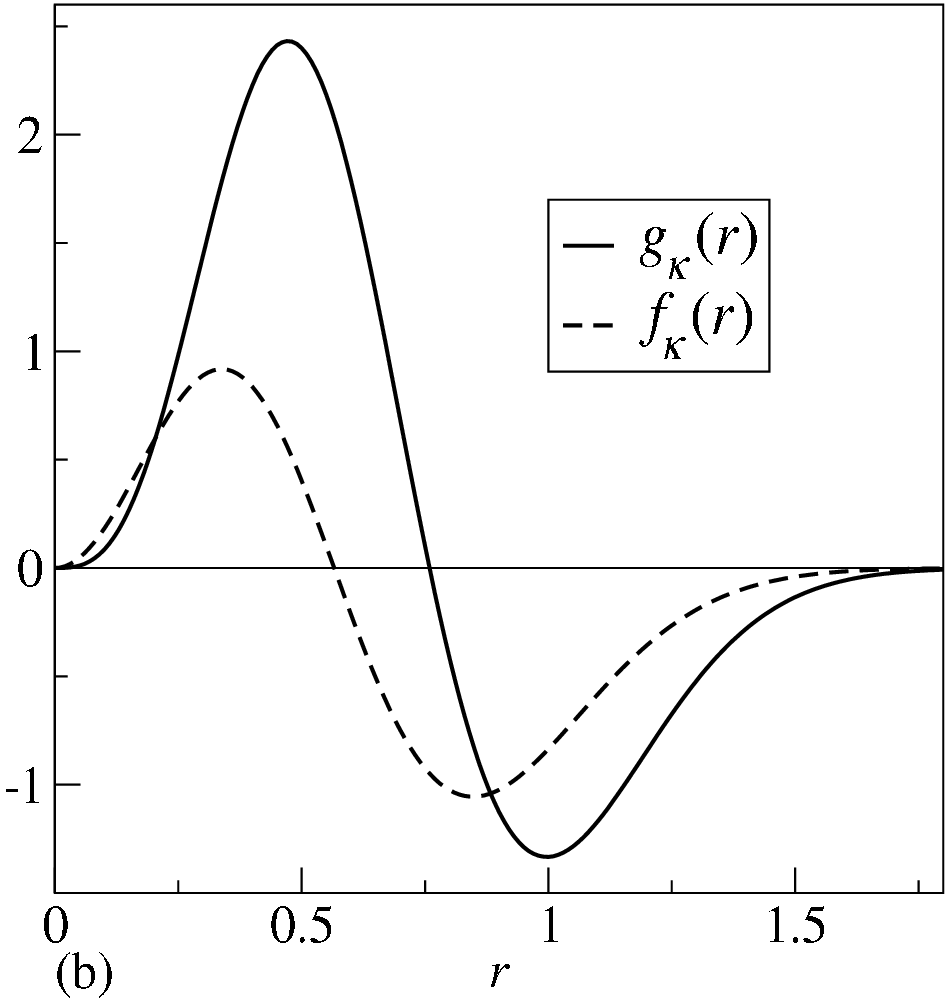}
\end{center}
}\hfill
\parbox[!ht]{5.5cm}{
\begin{center}
\includegraphics[width=5.5cm,height=5.5cm]{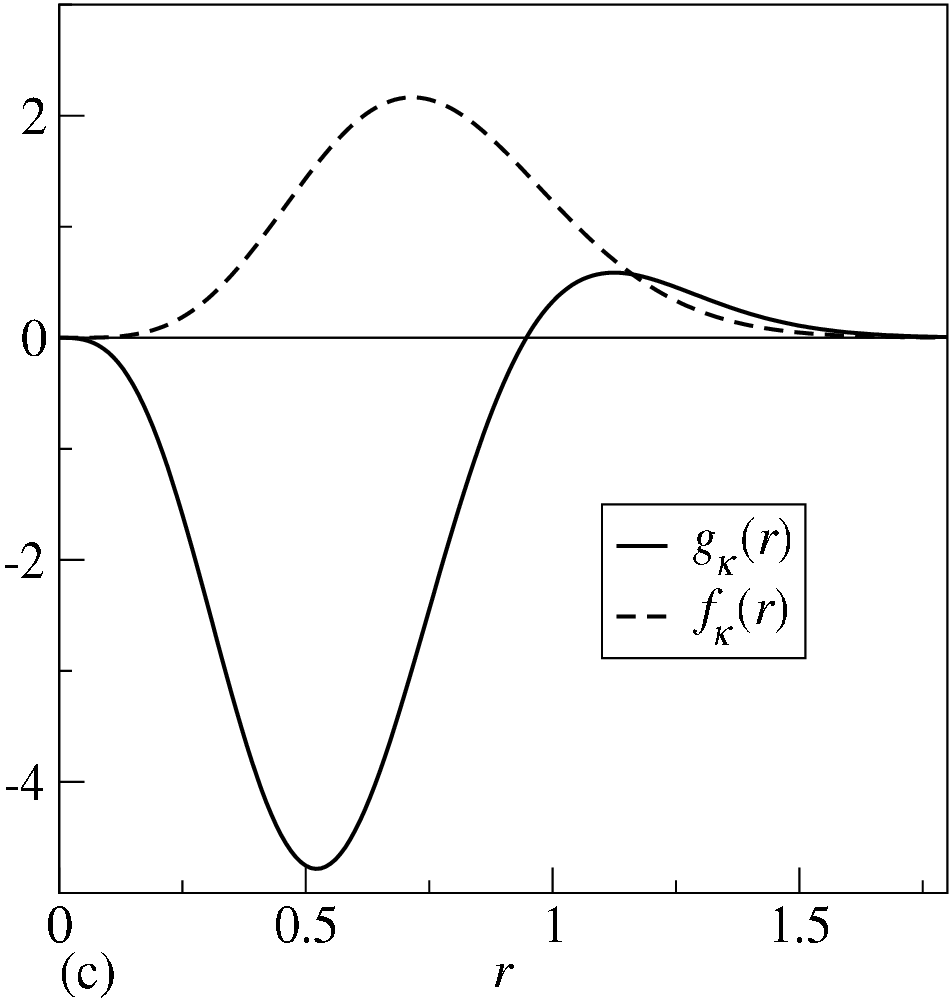}
\end{center}
} \vspace*{-0.2cm} \caption{Radial wave functions for the
potential $\Sigma=0$ of the states (a) $\tilde 1\tilde p_{1/2}$
($2s_{1/2}$), (b) $\tilde 1\tilde p_{3/2}$ ($1d_{3/2}$), and  (c)
$\tilde 0\tilde f_{5/2}$ ($1d_{5/2}$) with the same parameters as
in Fig.~\ref{Fig:spcPoDel}.\hfill\ } \label{Fig:PoSig}
\end{figure}

 These radial wave functions are represented in Fig.~\ref{Fig:PoSig} for
the $2s_{1/2}$, $1d_{3/2}$, and $2d_{5/2}$ states. Again, we see
that the node structure is the same as the one given by
Eq.~(\ref{Eq:nodes_Sigma0}). For the states $2s_{1/2}$ and
$1d_{3/2}$, which are pseudospin doublets as we will see later,
the lower components are equal. This is a general feature of
pseudospin symmetry, since when one acts upon a given state with
the pseudospin SU(2) generators to obtain its pseudospin
partner, the radial lower component is not changed because of the
particular form of these generators, as was shown in Refs.~\cite{levi,levi1}.

The eigenenergies for $\Sigma=U=0$ are obtained taking the limit $%
\omega _{2}\rightarrow \,0$ in Eq.~(\ref{Eq:AutS0}), which leads to
\begin{equation}
(\mathcal{E}+m)\sqrt{\frac{\mathcal{E}-m}{2m}}=\omega _{1}\left( 2\tilde{n}+%
\tilde{l}+\frac{3}{2}\right) \qquad (\tilde{n}=0,1,2,\ldots ).
\label{Eq:EnrOHpapertilde}
\end{equation}

Again, since the right-hand side of Eq.~(\ref{Eq:EnrOHpapertilde})
is positive and real, we see that the real solutions must have
positive binding energy $E=\mathcal{E}-m$. Using similar arguments
as with Eq.~(\ref{Eq:EnrOHpaper}), one can show that there can be
just one of such solutions.

\begin{figure}[!ht]
\parbox[!ht]{9cm}{
\begin{center}
\includegraphics[width=9cm]{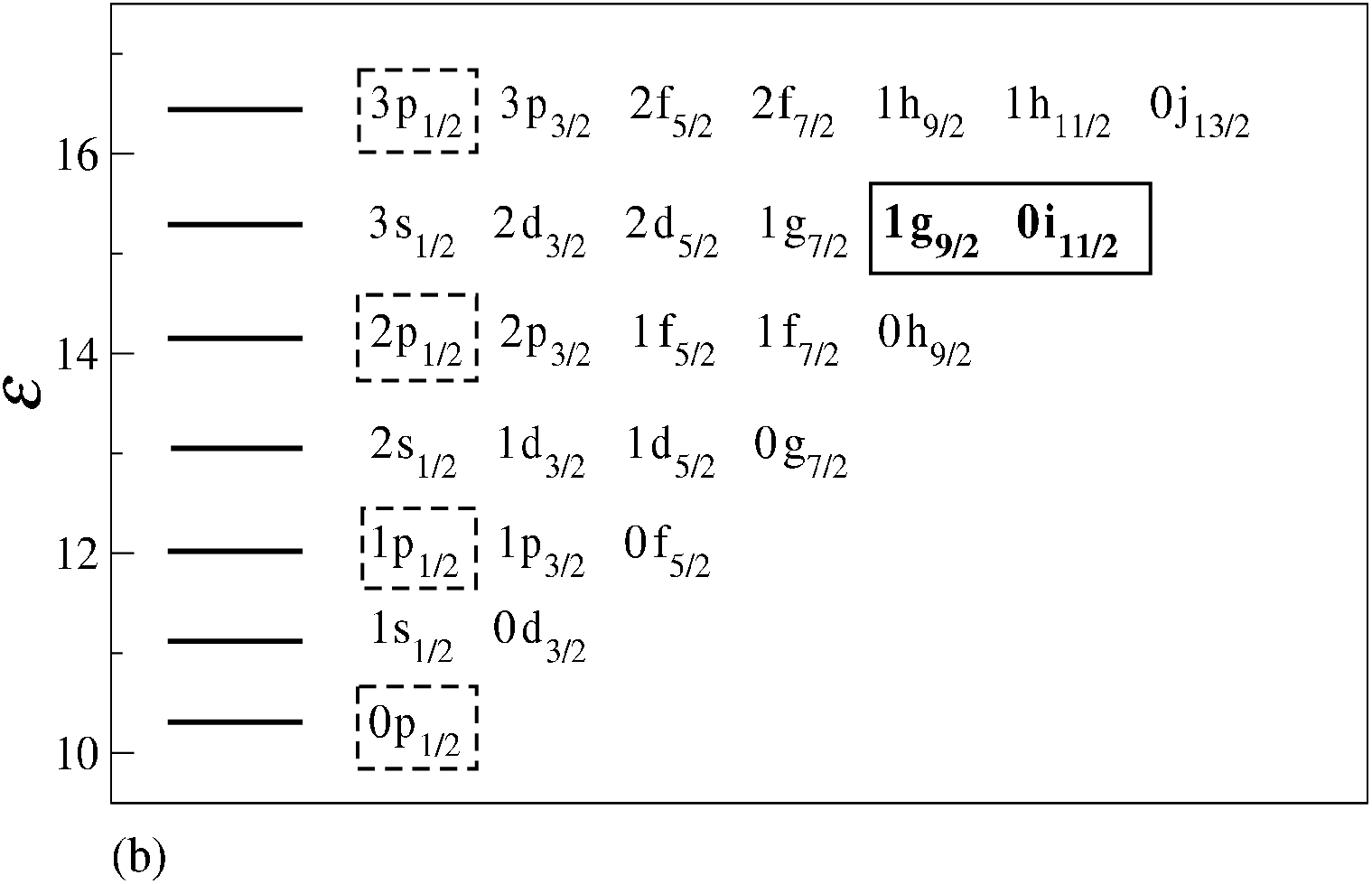}
\end{center}
}\vspace*{.5cm}
\parbox[!ht]{9cm}{
\begin{center}
\includegraphics[width=9cm]{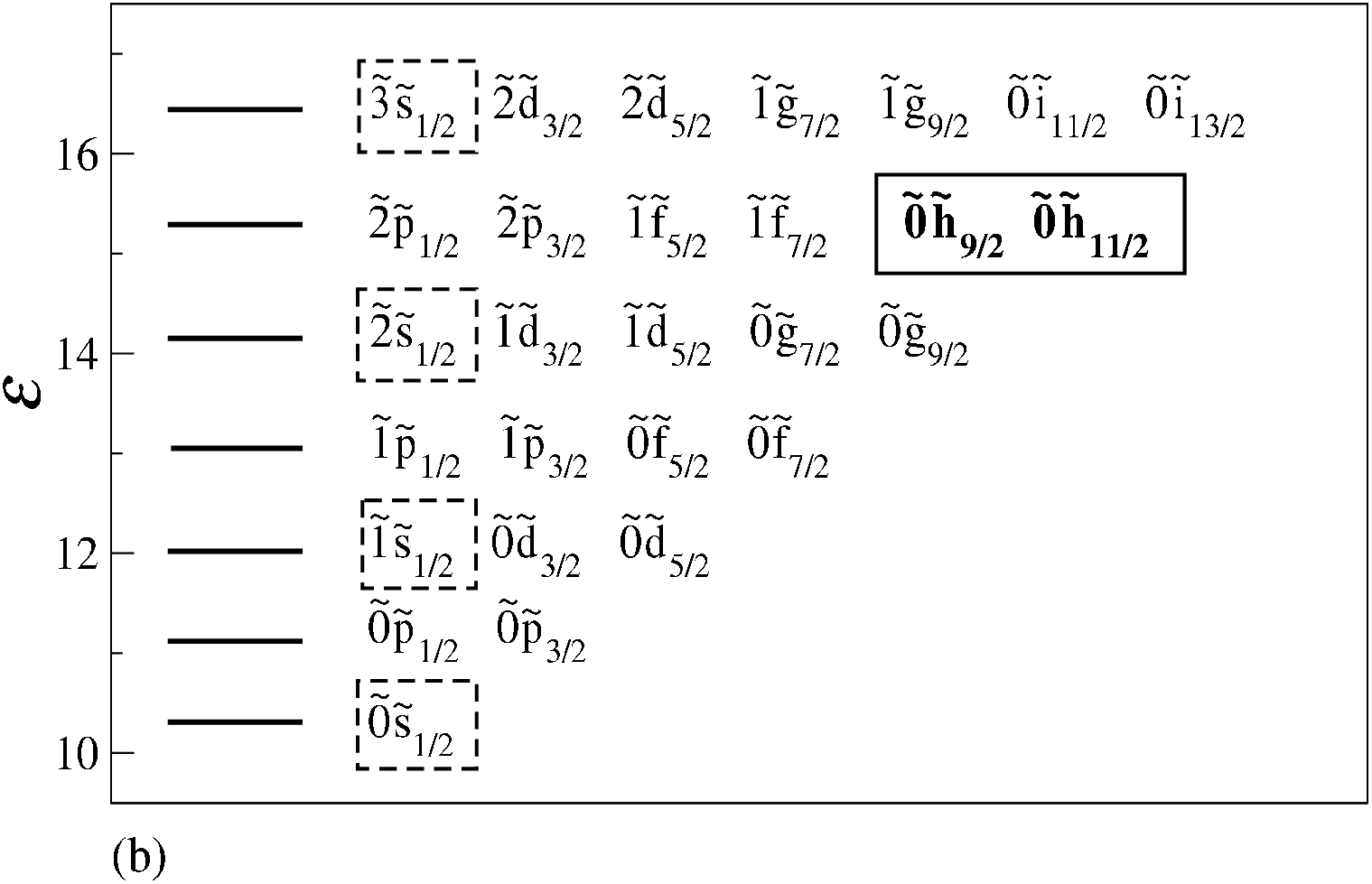}
\end{center}
}\hfill \caption{Single particle energies for the case $\Sigma=0$
with the quantum numbers (a) $(n\,l\,j)$ and (b)
$(\tilde{n}\,\tilde{l}\,{\tilde{j}=j})$. The parameters are
$\omega_{1}=2$, $\omega_{2}=0$, and $m=10$.\hfill.}
\label{Fig:spcPoSignl}
\end{figure}

The nonrelativistic limit in this case is obtained by setting $%
\omega _{2}=0$ in Eq.~(\ref{Eq:E-non-rel-Sigma_0}). As remarked
above, the resulting expression is of second order in $\omega
_{1}/m$, meaning that the energy is zero up to first order in
$\omega _{1}/m$. We can interpret this fact
by saying that, up to this order, there is no nonrelativistic limit for $%
\Sigma =U=0$ and therefore the theory is intrinsically
relativistic and so is the pseudospin symmetry. In this case, the
second order equation in Eq.~(\ref{Eq:D2ordfOFr2}), which only
depends on $\tilde{l}$, and also the eigenenergies in
Eq.~(\ref{Eq:EnrOHpapertilde}), show that there is no
pseudospin-orbit coupling and therefore the
states with same  $(\tilde{n},\tilde{l})$, but with $j=\tilde{l}+1/2$ and $j=%
\tilde{l}-1/2$, are degenerate. Thus, when $\Delta$ is a 
harmonic-oscillator potential and $\Sigma=U=0$, there are only
positive-energy bound states and exact pseudospin symmetry,
i.e., $\Delta$ acts as a binding potential. This is an
interesting result in view of the fact that the pseudospin symmetry
obtained in the limit $\Sigma(r)\rightarrow 0$ cannot be realized
for nuclear vector and scalar mean fields which go to zero as
$r\to\infty$, since in that case $\Sigma$ acts as a binding
negative central potential well and therefore no bound states may
exist when $\Sigma=0$ \cite{gino,pmmdm}.
The spectrum of single particle states for the case $\Sigma=0$ and $%
m=10$ is shown in Fig.~\ref{Fig:spcPoSignl}(a) using the quantum
numbers of the upper component, which can be seen as the analog
of the nonrelativistic quantum numbers. In
Fig.~\ref{Fig:spcPoSignl}(b) we classify the same energy levels by
the quantum numbers of the lower components $f_{\kappa}$. The
comparison between these two figures manifests the pseudospin
symmetry and its quantum numbers. For example, the doublets
$[1s_{1/2}-0d_{3/2}]$ and $[1p_{3/2}-0f_{5/2}]$, which have the
same pseudo angular momentum $\tilde{l}$ and the same $\tilde{n}$,
are, in the new notation, $[\tilde 0\tilde{p}_{1/2}-\tilde
0\tilde{p}_{3/2}]$ and $[\tilde 0\tilde{d}_{3/2}-\tilde
0\tilde{d}_{5/2}]$, respectively. Therefore,
the harmonic oscillator with $\Sigma=U=0$ and $\Delta=\frac{1%
}{2}m\omega _{1}^{2}r^{2}$ provides an example of exact pseudospin
symmetry. The fact that for this case we do not have a
nonrelativistic limit, already shows the relativistic nature of
the pseudospin symmetry. Figure \ref{Fig:spcPoSignl}(a) shows the
$(2\tilde{n}+\tilde{l})$ degeneracy, which means that not only
states with same $\tilde{n},\tilde{l}$ are degenerate (pseudospin
partners) but also, for example, $(\tilde{n}-1,\tilde{l}+2)$ or
$(\tilde{n}+1,\tilde{l}-2)$ have the same energy.

 The singlet states ($\tilde{l}=0$) are the states $np_{1/2}$ with
$l=1$ inside the small dashed squares in Fig. \ref{Fig:wfDel}. It
is interesting to point out that the ground state level for this
case is the state $0p_{1/2}$ with $l=1$. This is an unusual
behavior coming from the relativistic nature of this particular
case.


\section{\protect The pseudospin symmetry  and the intruder levels}

\label{PSymmetry}

The concept of pseudospin symmetry appeared in nuclear physics
more than 20 years ago \cite{npa137,plb30}. It was introduced to
account for the observation, in heavy nuclei, of the
quasidegeneracy of orbitals with quantum numbers (for fixed $n$
and $l$)
\begin{equation}  \label{Eq:nradial}
\left(n,\,l,\,j=l+\frac{1}{2}\right)\qquad\mathrm{and}\qquad
\left(n-1,\,l+2,\,j=l+\frac{3}{2}\right).
\end{equation}
Such pairs of single particle states are known as pseudospin
partners.

The doublet structure in the spectrum is better expressed in terms
of the pseudo-orbital angular momentum and pseudospin quantum number $\tilde{l}%
=l+1$ and $\tilde{s}=s=1/2$, respectively. The former, as noted by
Ginocchio \cite{gino}, is just the orbital angular momentum of the
lower component of the Dirac spinor, introduced in
Sec.~\ref{Sec:OscGeral}.
The pseudospin partners have the same $\tilde{l}$. For example, for the partners $%
[ns_{1/2},(n-1)d_{3/2}]$, $\tilde{l}=1$, and for
$[np_{3/2},(n-1)f_{5/2}]$ one has $\tilde{l}=2$. Yet, the total
angular momentum quantum number is still given by $j=\tilde{l}\pm
\tilde{s}$, where the aligned ($+$) and
unaligned ($-$) cases are determined by the sign of $\tilde{\kappa}$ [see Eq.~(%
\ref{al_unal_pspin})]. Pseudospin partners have $\tilde{\kappa}$
with opposite signs (the states with $n$ and $n-1$ radial quantum
numbers have, respectively, $\tilde{\kappa}>0$ and
$\tilde{\kappa}<0$) and thus pseudospin and pseudo-orbital angular
momentum have opposite alignments. The fact that $\tilde{l}$ is a
good quantum number can then be related to the disappearance of
the term with $\Sigma ^{\prime }$  in
Eq.~(\ref{Eq:D2ordfOHGeral}), which can be interpreted as
pseudospin-orbit interaction term \cite{prc58}. From that equation
it is also clear that this happens only when $U=0$, that is, the
tensor interaction breaks pseudospin symmetry.

The existence of pseudospin partners is connected to a SU(2)
symmetry of the Dirac equation with only scalar and vector potentials when $%
\Delta=0$ or $\Sigma=0$, rather than to the particular shapes of these
potentials. Back in 1975, Bell and Ruegg \cite{bell} obtained the pseudospin
generators for that symmetry. In that pioneering work only the case $%
\Delta=0 $ was studied with the possible application to  meson
spectra exhibiting a tiny spin-orbit splitting. More recently,
Ginocchio considered the other case, $\Sigma=0$, to explain the
quasi-degeneracy of pseudospin doublets in nuclei. The generators
of the symmetry for radial potentials were worked out in
Ref.~\cite{levi}.

It is interesting to discuss how the pseudospin symmetry gets
broken. The SU(2) generators of the pseudospin symmetry are given by \cite
{levi,smith,bell}
\begin{equation}
S_i=s_i\,\frac{1}{2}(1-\beta)+\frac{\mbox{\boldmath $\alpha$}\cdot%
\mbox{\boldmath $p$}\ s_i \,\mbox{\boldmath $\alpha$}\cdot%
\mbox{\boldmath
$p$}}{p^2}\,\frac{1}{2}(1+\beta)= \left(
\begin{array}{cc}
\tilde s_i & 0 \\
0 & s_i
\end{array}
\right)\,,
\end{equation}
where
\begin{equation}
\tilde s_i=\frac{\mbox{\boldmath $\sigma$}\cdot\mbox{\boldmath $p$}}{p} s_i
\frac{\mbox{\boldmath $\sigma$}\cdot\mbox{\boldmath $p$}}{p}=\frac{2%
\mbox{\boldmath $s$}\cdot\mbox{\boldmath $p$}}{p^2} p_i-s_i,
\end{equation}
and $s_i={\sigma_i/2}$ ($i=1,2,3$). The commutator of this
operator with the Hamiltonian in Eq.~(\ref{1}) with $U=0$ is
\begin{equation}
[H_{\mathrm{D}},S_i]=\left(
\begin{array}{cc}
[\Sigma,\tilde s_i] & 0 \\
0 & 0
\end{array}
\right)\, .  \label{comut}
\end{equation}
Thus, the pseudospin symmetry breaking can be related to the
commutator $[\Sigma,\tilde s_i]$ \cite{dudek}. This commutator is
zero when $\Sigma=0$, but for radial potentials it is enough that
$\Sigma^{\prime}(r)=0$ \cite{prc58}. The quasidegeneracy of some
pseudospin partners can be seen in a Dirac Hamiltonian with scalar
and vector potentials of Woods-Saxon type \cite{pmmdm}. As
explained before, one cannot set $\Sigma=0$ for those kind of
potentials, so we cannot get the full degeneracy of all pseudospin
partners. In fact, the nuclear pseudospin symmetry in nuclei has a
dynamical character coming from a cancellation among different
terms that contribute to the energy as discussed in several works
\cite{pmmdm,ronai,marcos,marcos2}.

We have seen in the previous sections that the relation between
the nodes of radial functions depends on whether one has $\Delta
=0$ or $\Sigma =0$ in Dirac Hamiltonian with harmonic-oscillator
potentials. Leviatan and Ginocchio have shown \cite{gino2} that,
in the limit of pseudospin symmetry, the lower radial functions of
the pseudospin partners have the same number of nodes. Using this
information, they were able to show \cite{levi1} that, for scalar
and vector potentials which go to zero at infinity, like
Woods-Saxon potentials, the number of nodes is related by
\begin{eqnarray}
n_{f}=\left\{
\begin{array}{ll}
n_{g}\,, &\;\ \ \ \kappa<0\\[0.2cm]
n_{g}+1\,,   &\;\ \ \ \kappa>0.
\end{array}
\right.\label{Eq:nodes}
\end{eqnarray}

This is the behavior we have found before for scalar and vector
oscillator harmonic potentials with $\Delta =0$,
Eq.~(\ref{Eq:nodes_Delta0}), precisely the case for which we have
the nonrelativistic harmonic-oscillator limit.

Let us now discuss the levels with $n_{g} = n_{f}=0$. From
Eq.~(\ref{Eq:nodes}), we see that this can only happen when
$j=l+1/2 \, (\kappa<0)$ but never with $j=l-1/2 \, (\kappa>0)$. If
we recall that the pseudospin doublets have the same $n_f$, in
this case $n_{f}=0$, states such as
$0s_{1/2},\,0p_{3/2},\,0d_{5/2}$, etc., have no pseudospin partners.
These states are known as \emph{intruder} states. The $0g_{9/2}$
state, for example, is an intruder and what would be its
pseudospin partner
--- the state $0i_{11/2}$ ($\kappa =6$) --- does not exist
because $n_{g}=n_{f}-1$ would be negative for $n_{f}=0$. This is
what we observe in the nuclear spectra.

In the case of exact pseudospin symmetry ($\Sigma=0$) with a
harmonic oscillator potential for $\Delta$, which we have
discussed in Sec.~\ref{SubSec:omega2_0_S_0}, the structure of the nodes
it is quite different. In accordance with
Eq.~(\ref{Eq:nodes_Sigma0}),
\begin{eqnarray}
n_{f}=\left\{
\begin{array}{ll}
n_{g}-1\,, &\;\ \ \ \kappa<0\\[0.2cm]
n_{g}\,,   &\;\ \ \ \kappa>0.
\end{array}
\right.\label{Eq:nodes1_Sigma0}
\end{eqnarray}

Thus the intruder levels that have $n_{g} =n_{f}=0$ may only exist
in the exact pseudospin limit when $j=l-1/2 \, (\kappa>0)$, as we
see from Eq.~(\ref{Eq:nodes1_Sigma0}) and, in that sense, they are
not the intruder levels seen in nuclei, where $j=l+1/2 \,
(\kappa<0)$.

In a recent paper the intruder levels have been discussed in the
context of the relativistic harmonic oscillator \cite{Chen_Meng}
and it was suggested that they will have partners in the limit of
the exact pseudospin symmetry ($\Sigma=0$). However, in that
paper, the pseudospin partners have been classified considering
only the radial quantum number of the lower component (same
$n_{f}$). In that classification, levels with the same $n_{f}=0$
may have partners according to Eq.~(\ref{Eq:nodes1_Sigma0}).
However, it seems that the relation between the number of nodes of
the upper and lower components was considered to be given by
Eq.~(\ref{Eq:nodes}) instead of the correct one given by
Eq.~(\ref{Eq:nodes1_Sigma0}).
 If we take into account that the nuclear intruder levels have $j=l+1/2
 \ (\kappa<0)$ then, by Eq.~(\ref{Eq:nodes1_Sigma0}), we see that
states with negative $\kappa$ that have zero nodes in the lower
component will have a node in the upper component, $n_{g}=1$. This
disagrees with nuclear spectroscopy, since the intruder states
should have $n_{g}=0$. This is shown in Fig.~\ref{Fig:spcPoSignl}(a) 
where we show inside the solid square the state $1g_{9/2}$ with
$n=1$ and its partner $0i_{11/2}$ that in the
($\tilde{n},\tilde{l}$) classification in
Fig.~\ref{Fig:spcPoSignl}(b) are the pseudospin doublet [$\tilde
0\tilde{h}_{9/2} - \tilde 0\tilde{h}_{11/2}$].

We can summarize this discussion and conclude that even though the
harmonic oscillator with $\Sigma=0$ has an exact pseudospin
symmetry that could explain the almost degeneracy in $\tilde l$
observed in nuclei for the doublets with $j=l+1/2 \, (\kappa<0)$,
$j=l-1/2 \, (\kappa>0)$, radial and orbital quantum numbers
related as shown in Eq.~(\ref{Eq:nradial}), it cannot explain the
origin of the intruder levels. Furthermore, the harmonic
oscillator presents a much higher degeneracy in the energy levels
besides the simple degeneracy $\tilde l$, as we have shown, which
implies that the nuclear potential cannot be approximated by a
harmonic oscillator and must be more complicated in order to
remove this extra degeneracy. The asymptotic properties of the
central mean-field potential also seem to play a crucial role in
defining the node structure of the single particle levels.

\section{\protect Conclusions}

\label{conclusions}

We have presented a generalized harmonic oscillator for spin 1/2
particles that includes not only the usual linear tensor potential
obtained by a redefinition of the momenta, but also a vector $(V)$
and scalar potential $(S)$ that appear in the Dirac Hamiltonian in
the specific combinations $1/2(1\pm\beta)V$, in which $V$ is a
harmonic oscillator radial potential. A special attention was paid
to the $1/2(1-\beta)V$ combination, i.e., when $S=-V
(\Sigma=0)$, which has been related recently to the nuclear
pseudospin symmetry.

We have derived the energy eigenvalue equations and shown explicitly 
the wave functions. The
nonrelativistic limits for all  cases have been discussed. The
analytical expression for the upper and lower components of the
Dirac spinor for $U = 0$, when $\Sigma=0$ (or $\Delta= 0$), has
been found.

We have discussed the structure of the solutions of the eigenvalue
equations and presented explicitly the positive energy solutions.
We have concluded that only when the tensor potential is turned on
we have the negative bound state solutions and in this case, when
$\omega_1=0$, the two bound state solutions are symmetric. When
the tensor potential is absent we have also shown that the special
conditions between the scalar and vector potential ($S=V$ or
$S=-V$) needed to have an harmonic-oscillator potential with
scalar and vector potentials exclude the negative bound state
solutions from the spectra.

 The structure of the radial nodes for the Dirac spinor has also been
presented and compared to the case when the potentials vanish for
large distances. We discussed in detail the case of the harmonic
oscillator with $U=\Sigma=0$ for which pseudospin symmetry is
exact, i.e., pseudospin doublets (states with same
$\tilde{l}$) are degenerate. In this case we can have bound states
due to the positive harmonic-oscillator potential $\Delta$. These
states can be regarded as intrinsically relativistic, in the sense
that the $\Delta$ potential has no analog in the
nonrelativistic limit. In fact, they show a very peculiar
structure of radial nodes in comparison to the nonrelativistic
harmonic-oscillator radial wave function: the states with $j = l +
1/2$ have at least one node $(n \geq 1)$ in the upper component
and can have zero nodes only in the lower one. As a consequence,
the intruder levels on the nuclear spectra (isolated levels with
$n=0$ and $j = l + 1/2$) do not exist in the harmonic oscillator
in the limit of nuclear pseudospin symmetry $(\Sigma = 0)$. In
this limit the states with zero nodes in the upper component have
$j = l - 1/2$. Moreover, the relativistic harmonic oscillator in
the case $U=\Sigma=0$ has a much higher energy degeneracy than the
approximate pseudospin symmetry seen in the nuclear spectra
(doublets with same $\tilde{n},\tilde l$). This is not surprising
since the meson exchange theory of nuclear forces produces nuclear
potentials that vanish when $r\rightarrow\infty$, a completely
different asymptotic behavior from the harmonic-oscillator
potential discussed here. One possible way to break this extra
degeneracy of the harmonic oscillator is to add a Woods-Saxon-like
potential as it has been done in Ref.~\cite{Chen_Meng}.

Another important conclusion from our work is that a Dirac
equation with scalar and vector harmonic-oscillator potentials
with the same sign and magnitude ($\Delta=0$), instead of the
linear tensor potential $U$, is the most natural way to introduce
the harmonic oscillator in relativistic quantum mechanics, since
one gets the correct nonrelativistic limit, so that the energy
levels and the upper component of the wave functions are very
similar to the non-relativistic ones.

 Finally, our analysis of the relativistic harmonic oscillator shows,
 in a very simple and analytical way, that the nuclear pseudospin
symmetry $(\Sigma=0)$ does not have a nonrelativistic limit, in
the sense that the eigenvalues are of second order in a $\omega /
m$ expansion. So we conclude that the relativistic harmonic
oscillator is an excellent example to show explicitly what
Ginocchio had already emphasized in his pioneering work: the
pseudospin symmetry in the Dirac Hamiltonian (when $\Sigma = 0$)
is indeed a symmetry of relativistic nature.

\vskip0.5cm We acknowledge financial support from CNPq, FAPESP, and
FCT (POCTI) scientific program. R.L. and M.M acknowledge, in
particular, the CNPq support and A.S.C was also supported by
FAPESP. P.A. and M.F. were supported in part by FCT (Lisbon),
Project No.~POCTI/FIS/451/1994.


\end{document}